\documentclass[10pt, letterpaper]{sig-alternate-10pt}

\usepackage{graphics}
\usepackage{verbatim}
\usepackage{epsfig}
\usepackage{latexsym}
\usepackage{graphicx}
\usepackage{balance}
\usepackage{subfigure}
\usepackage{algorithm}
\usepackage{algorithmic}
\usepackage{verbatim}
\usepackage{graphicx}
\usepackage{array}
\usepackage{bm}
\usepackage{color}
\usepackage{amsmath}
\usepackage{color}
\usepackage{graphicx}
\usepackage{enumitem}
\usepackage{lipsum}
\usepackage{epstopdf}
\usepackage{url}
\usepackage{makecell}
\begin{document}
	

	
	
	
	
	%
	
	\title{EchoLock: Towards Low Effort Mobile User Identification}
	%
	%
	%
	%
	%
	
	\numberofauthors{4} 
	%
	\author{
	%
	%
	\alignauthor
	Yilin Yang\\
	\affaddr{Rutgers University}\\
	\email{yy450@scarletmail. \linebreak[0] rutgers.edu}
	\alignauthor
	Chen Wang\\
	\affaddr{Louisiana State University}\\
	\email{chenwang1@lsu.edu}
	\thanks{This work was done during Chen Wang's Ph.D. study at Rutgers University.}
	\alignauthor Yingying Chen\\
	\affaddr{Rutgers University}\\
	\email{yingche@scarletmail. \linebreak[0] rutgers.edu}
			\and
	\alignauthor Yan Wang\\
	\affaddr{Binghamton University}\\
	\email{yanwang@binghamton.edu}
	}
	
	\maketitle
	
	\global\csname @topnum\endcsname 0
	\global\csname @botnum\endcsname 0
	
	\begin{abstract}

User identification plays a pivotal role in how we interact with our mobile devices. Many existing authentication approaches require active input from the user or specialized sensing hardware, and studies on mobile device usage show significant interest in less inconvenient procedures. In this paper, we propose \textit{EchoLock}, a low effort identification scheme that validates the user by sensing hand geometry via commodity microphones and speakers. These acoustic signals produce distinct structure-borne sound reflections when contacting the user's hand, which can be used to differentiate between different people based on how they hold their mobile devices. We process these reflections to derive unique acoustic features in both the time and frequency domain, which can effectively represent physiological and behavioral traits, such as hand contours, finger sizes, holding strength, and gesture. Furthermore, learning-based algorithms are developed to robustly identify the user under various environments and conditions. We conduct extensive experiments with 20 participants using different hardware setups in key use case scenarios and study various attack models to demonstrate the performance of our proposed system. Our results show that \textit{EchoLock} is capable of verifying users with over 90\% accuracy, without requiring any active input from the user. 

\end{abstract}

\section{Introduction}
\label{sec:intro}

\begin{figure}
	\centering
	\includegraphics[width=3in]{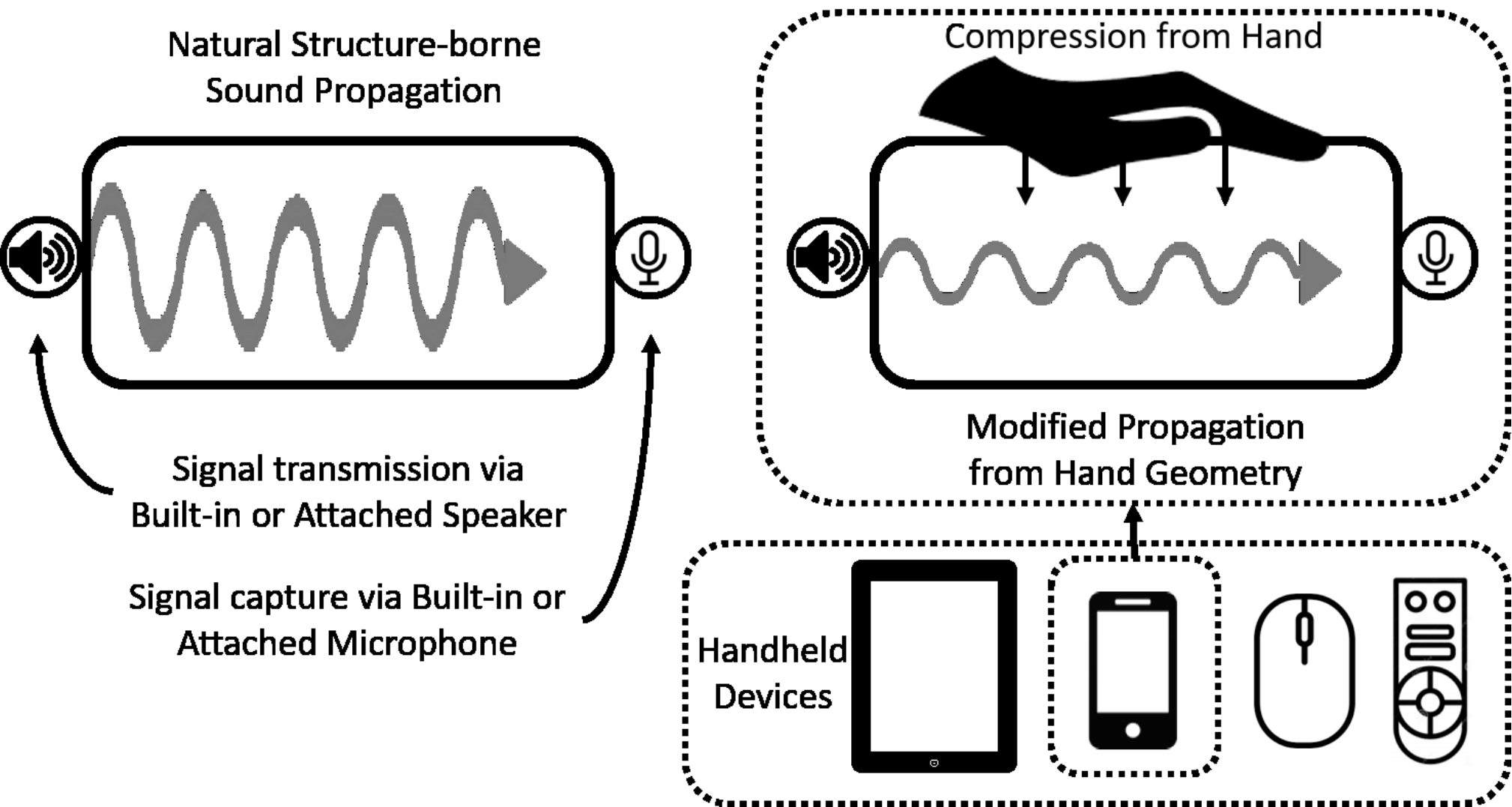}
	\vspace{-3mm}
	\caption{Capture of hand biometric information embedded in structure-borne sound using commodity microphones and speakers.}
	\vspace{-5mm}
	\label{fig:basic}
\end{figure}

User identification is a fundamental and pervasive aspect of modern mobile device usage, both as a means of maintaining security and personalized services. Verifying oneself is necessary to gain access to smartphones, bank accounts, and customized news feeds; information and resources which must be available on demand. As such, repeated acts of authentication can quickly grow tedious and consume unnecessarily long portions of daily routines involving mobile devices. Studies on cellphone addiction suggest that user identification procedures encompass up to 9.0\% of daily usage time \cite{harbach2014hardlock}, with related inquiries showing strong interest in more convenient practices \cite{vanbruggen201userbehave}. Techniques such as facial recognition or fingerprinting provide fast validation without requiring considerable effort from the user, but demand dedicated hardware components that may not be available on all devices. This is of particular importance in markets of developing countries, where devices such as the Huawei IDEOS must forgo multiple utilities in order to maintain affordable price points (e.g. under \$80) \cite{ibtimes2017, singularity2011}. Secure and effective functionality necessitates a lightweight protocol to distinguish between different users, both to facilitate intelligently tailored services as well as maintain privacy.

To this end, we propose \textit{EchoLock}, an original and inexpensive technique capable of secure, fast, and low-effort user identification utilizing only commodity speakers and microphones. \textit{EchoLock} uses a carefully designed inaudible signal to measure acoustic reflections and structural vibrations produced by the user's hand when holding their mobile devices, such as smartphones, tablets, smart remotes, or wireless mouses. Sound waves produced by speakers propagate through physical mediums such as the aforementioned devices in the form of structure-borne sound. Pressure from the hand applied to the device has a unique and observable impact on structure-borne sound propagation. This information is representative of the user's hand geometry, a biometric indicator well studied in medical fields and known to be accurate \cite{barra2019handbio}, yet rarely employed in mobile applications due to obstacles in obtaining accurate measurements with limited hardware. We find that acoustic sensing of these characteristics is not only feasible, but also low-effort as no conscious action is required by the user; holding the device itself is the validating action, rather than typing a password or pressing a fingerprint scanner, as shown in Figure~\ref{fig:basic}. This can be achieved using only speakers and microphones found on most commercial-off-the-shelf (COTS) devices, making our system non-intrusive and low cost. The availability of these hardware components is only expected to increase with the rising prevalence of integrated IoT devices built with virtual assistants and voice controllers, projected to reach an install base of over 75 billion by 2025 ~\cite{statista2019}.

Existing solutions in the market are typically considered effective and fast, but do have some limitations regarding ease of use. Actions such as password entry, voice utterances, or finger presses demand, however briefly, the user's attention and active participation in the process. In contrast, \textit{EchoLock} is a passive procedure. Our technique serves as a viable standalone identification system, or as a complementary system for multi-factor authentication due to its naturally low involvement. Password security, for example, can be enhanced by simultaneously sampling hand geometry during typing or swiping actions, compensating for common vulnerabilities (e.g. PIN codes spied on through shoulder-surfing attacks cannot be used if the attacker's hand is not recognized by the device).

As a standalone technique, \textit{EchoLock} is applicable to a wide variety of services. Resources such as financial accounts or health apps on smartphones can trigger a single-instance identification check to verify the user's hand before divulging sensitive information. Private message notifications can be displayed or hidden onscreen depending on which person is currently holding the device. The low costs of small-scale microphone and speaker sensors can enable even unconventional devices to be compatible with \textit{EchoLock}. Door handles augmented with such components can passively sense structure-borne sound propagation as the user holds them and open or lock accordingly. The speed of sound propagation is rapid, even when traveling through physical mediums, enabling our identification checks to be completed within milliseconds. Continuous authentication can also be implemented via periodic measurements.

Building \textit{EchoLock} for such applications does present many challenges, however. The most prominent is the challenge in leveraging a single pair of low-fidelity speaker and microphones to capture the unique characteristics of a user's hand geometry, which usually has only minute differences between people.
In addition, the acoustic signal propagating from the device's speaker to its microphone usually experiences multipath effect, resulting in airborne and structure-borne signals that are hard to distinguish from each other.
Moreover, the ambient noises and acoustic signals reflected off surrounding obstacles create serious interference that needs to be accounted for. Last but not least, many factors could impact the robustness of the proposed approach, such as different device shapes or materials.

To address these challenges, \textit{EchoLock} utilizes an ultrasonic signal to sense a user's mannerisms when holding a device. A high-frequency, short duration transmission is selected to reduce audible disturbances to the user and provide prompt validation. The signals coming through the structure-born and near-surface airborne propagation are separated based on different sound speeds in the air and solid materials~\cite{material2019}. The system applies a band-pass filter to remove the ambient acoustic noises that do not share the same spectrum as the designated ultrasonic signal. We derive fine-grained acoustic features, including statistic features in the time and frequency domain as well as MFCC features, to capture the unique hand holding biometrics. We find simple learning-based algorithms (i.e., SVM and LDA) are sufficient to robustly identify the user considering various impact factors. There is no dependence on specialized cameras or other dedicated hardware to accomplish this, requiring only the use of readily accessible microphones and speakers found on most modern mobile devices. Most significantly, no active input is required from the user, making our system low effort.

Our main contributions in this work are as follows:
\vspace{-2mm}
\begin{itemize}
	\item We develop a low-effort user identification system for mobile and smart handheld devices (e.g. smartphones, smart remotes, wireless mouses) that validates hand biometric information based on acoustic sensing. The proposed system does not require any input from the user and is non-invasive by utilizing a inaudible frequencies. Moreover, our minimal hardware requirements make our system easy to be deployed on most COTS devices.
\vspace{-2mm}
	\item We design an acoustic sensing-based approach to study the unique signal interferences caused by an individual's hand on sound propagation resulting from their different curvatures, finger sizes and holding behavior. We demonstrate that structure-born sound propagation of a carefully designed acoustic signal is able to carry physiological and behavioral hand information, which previously relied on imaging techniques to measure.
\vspace{-2mm}
	\item We identify unique acoustic features, including time-domain, frequency domain, and MFCC features, to capture the user's biometrics and apply signal processing and learning-based methods to distinguish the users based on their phone-holding behaviors for user identification.
\vspace{-2mm}
	\item We implemented an early prototype of \textit{EchoLock} on various mobile devices and evaluated performance under multiple conditions. Our tests involving over 160 trials of key use case scenarios show identification accuracy upwards of 90\%.
\end{itemize}

\begin{table*}
	\footnotesize
	\small
	\caption {Qualitative comparison of existing user identification methods.} \label{tab:quality}
	\begin{center}
		\begin{tabular}{|c|c|c|c|c|c|c|c|c|}
			\hline
			\makecell{Identification \\ Technique} & 
			\makecell{Evaluation \\ Category} & 
			\makecell{Personally \\ Identifiable} & \makecell{Physiological \\ Credentials} & \makecell{Behavioral \\ Credentials} & 
			\makecell{Specialized \\ Hardware}\\ 
			\hline
			Lockscreen Image \cite{chiang2013password} & Knowledge & No & No & Yes & No 
			\\ \hline
			Face \cite{fathy2015face} & Visual & Yes & Yes & No & No 
			\\ \hline
			Fingerprint \cite{mathur2016fingerprint} & Visual & Yes & Yes & No & Yes
			\\ \hline
			Iris \cite{casanova2010iris} & Visual & Yes & Yes & No & Yes
			\\ \hline
			Gait \cite{zhong2014gait} & Visual & No & Yes & Yes & No
			\\ \hline
			Voice \cite{johnson2013voice} & Acoustic & Yes & Yes & Yes & No
			\\ \hline
			\textbf{Our Work (EchoLock)} & \textbf{Acoustic} & \textbf{No} & \textbf{Yes} & \textbf{Yes} & \textbf{No}
			\\ \hline
		\end{tabular}
	\end{center}
\end{table*}

	\vspace{-2mm}
\section{Related Work}
\label{sec:related}


Reconciling the trade-off of \textit{accuracy} and \textit{usability} is a constant challenge of user identification for mobile devices. 
The most popular identification methods are based on memorization of a text or numerical string (e.g., a password and PIN)~\cite{tsys2016us}.
For these methods, the user must either commit to memory complex knowledge (e.g., random characters or long combinations), or settle for a simpler string at the expense of security.
Graph-based authentication, such as lock patterns~\cite{uellenbeck2013quantifying} and image-based authentications~\cite{suo2005graphical, chiasson2007graphical}, are proposed to ease the mental burden. They identify a user by asking the user to swipe line segments across multiple grid points or recognize the user's pre-selected image from a sequence of randomly displayed pictures.
However, these approaches require more time to input graphical patterns compared to the entry of alphanumerical passwords. Moreover, all these knowledge-based credentials are subject to loss and have low security. 
To relieve the burdens of user input (e.g., memorization and input time) during identification, mobile devices tend to adopt biometric-based approaches, which allow the users to simply use physical traits of a body part as credentials.
Several popular examples include face ID~\cite{faceid2018}, capacitive fingerprint scanning~\cite{cherapau2015impact, mikael2015}, and iris scanning~\cite{Samsung2018}. These physical traits are typically unique to a person and do not change abruptly over time, making them ideal for identifying purposes. However, these approaches require expensive or dedicated hardware components to obtain users' physiological characteristics, which limits the deployment of these methods on most devices. Moreover, researchers show that voices or faces are faster to input than passwords and gestures, but users spend over $5$s on average to respond to the system with an active input in addition to the authentication processing time, which is not convenient~\cite{trewin2012biometric}. 

Human behaviors are considered another type of biometrics that can be used for identification. Because they usually encompass traits that are exhibited subconsciously, they are difficult to be stolen or imitated.
For example, existing works show that it is possible to identify a user based on finger gestures on the touch screen~\cite{napa2012biometricrich, shahzad2013secure,ren2015critical}, hand gestures in the air~\cite{kratz2014airauth}, finger input on any solid surface~\cite{liu2017vibwrite}, or voice commands~\cite{wechatvoiceprint, Siri}. 
While these approaches are low-cost, they also pose a challenge for identification due to a high level of variability when measuring these behaviors with low-fidelity sensors available on mobile devices. Moreover, behavioral-based identification still requires the user to actively input a behavior pattern, which is still not convenient for frequent device unlocking activities. 
There is active research on low-effort passive user authentication, which we consider to be the closest category of works to \textit{EchoLock}. 
Ren \textit{et al.} derive the unique gait patterns from the user's walking behavior to passively verify the user's identity by using accelerometer readings in mobile devices~\cite{ren2014user}. Zheng \textit{et al.} propose to extract the user's behavioral patterns of touchscreen tapping (e.g., rhythm, strength, and angle preferences of the applied force) from the device's built-in accelerometer, gyroscope, and touchscreen sensor to provide non-intrusive user authentication~\cite{zheng2014you} when the device is in the unlocked status. 

We summarize these findings in Table \ref{tab:quality}. Different from other approaches, \textit{EchoLock} utilizes novel hand-related physiological biometrics (e.g. hand contours, palm size, and finger dimensions) and behavioral biometrics (e.g. holding strength and behavior) to provide convenient and secure user authentication on mobile devices. In particular, we perform fine-grained acoustic sensing (i.e. using the near ultrasound emanations of the mobile device) to capture the unique characteristics of a user's holding hand for identification. 
Acoustic sensing is an emerging technique that has been widely used in many mobile computing applications (e.g., indoor localization~\cite{tung2015echotag}, computer-human interaction~\cite{tung2016forcephone,sun2018vskin}) because it is easy to obtain and process the acoustic signals in mobile devices. 
To our best knowledge, we are the first work to utilize acoustic sensing to capture biometric information for low effort user identification. Our proposed technique does not depend on personally identifiable information, active user inputs, long input time, or specialized hardware.

	\section{Feasibility Study}
\subsection{Sound Propagation on Mobile Devices}
 
 \begin{figure}[t] 
 	\centering
 	\includegraphics[width=\linewidth]{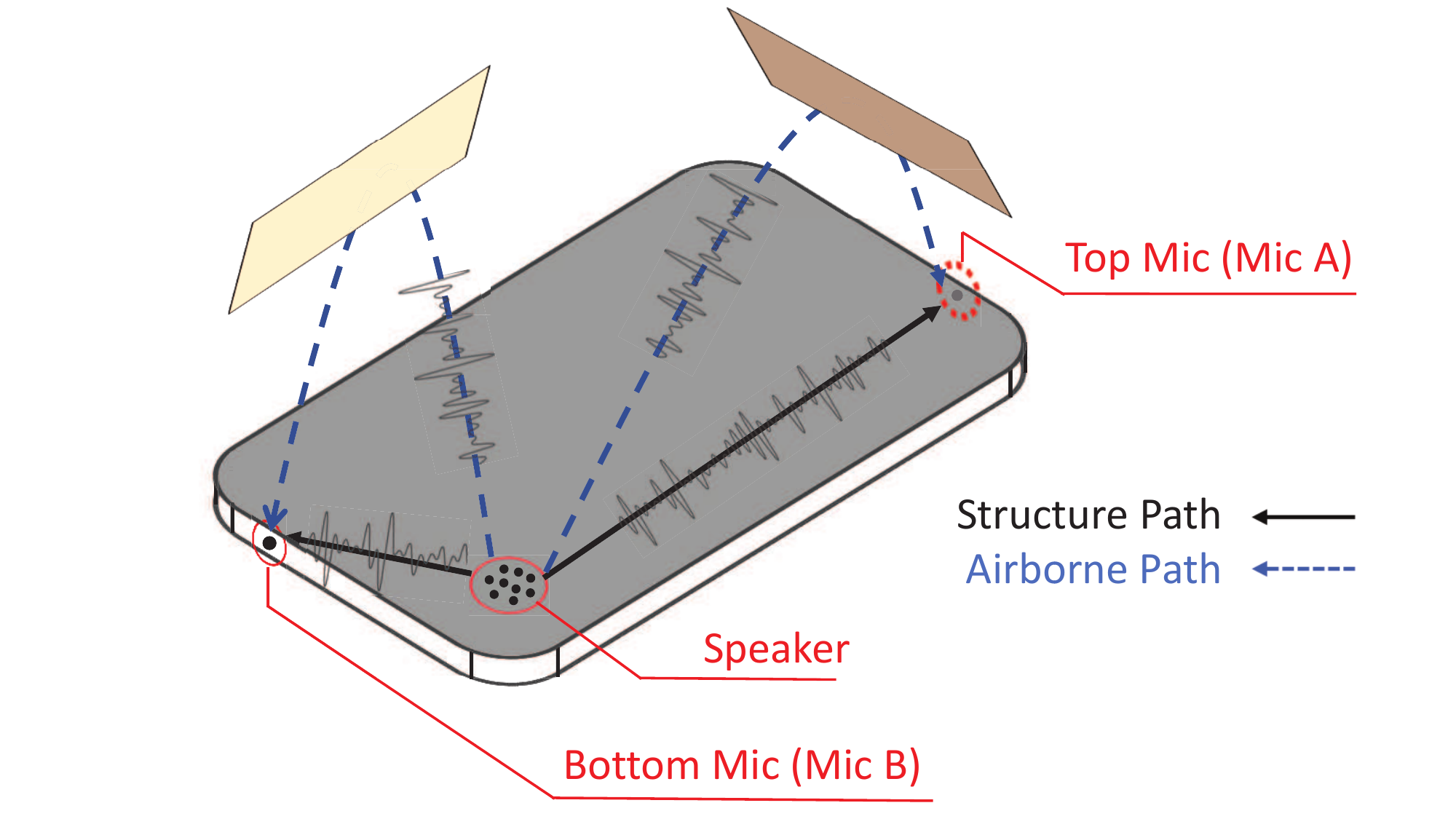}
 	\vspace{-6mm}
 	\caption{Example sensor configuration and sound paths for a typical smartphone design.}
 	\vspace{-3mm}
 	\label{fig:sound} 
 \end{figure}

Acoustic sensing techniques may vary depending on the transmission medium sound propagates through. Figure \ref{fig:sound} illustrates this using an example diagram of a common configuration format for speaker and microphone placement on a COTS smartphone. Many mobile devices are equipped with at least two microphones, positioned opposite of each other at either ends of the device. Sound sources, such as speakers of the mobile device, produce acoustic signals, which follow an airborne propagation path to be received by the microphones. 


%


For structure-borne propagation, however, the transmission medium is the device itself. This form of sound is most often recognized as vibration and can be perceived both aurally as well as tangibly, due to the nature of the medium. From Hooke's Law \cite{rychlewski1984hooke}, the speed of sound through a medium can be represented as a function, formulated as $c = \sqrt{\frac{K}{p}}$ where $K$ is the bulk modulus of elasticity, or Young's modulus, and $p$ is the medium density. As seen in Figure \ref{fig:sound}, the structure path is more direct due to the greater density and compression resistance of the mobile device, allowing sound to travel and be received more quickly compared to in the air. 
This trait is of interest as propagation through a physical medium provides natural resilience to reflections from distant obstacles as there is minimal deviation from the sound path.

A caveat to this, however, is that structure-borne propagation is much more sensitive to physical disturbances. Interactions such as touching the medium can significantly alter the acoustic patterns as the contact and force exerted upon the medium changes how it reverberates. While this normally poses a challenge for acute acoustic sensing, \textit{EchoLock} exploits this for the purposes of recognizing individual people. The force of a user's grip on the mobile device is integral to the system, essentially extending the medium to encompass both the device and user's hand. The bulk modulus of elasticity can be expressed as: 

\begin{figure}[t]
	\centering
	\subfigure[Held in hand]{%
		\label{fig:first}%
		\includegraphics[width=0.33\linewidth]{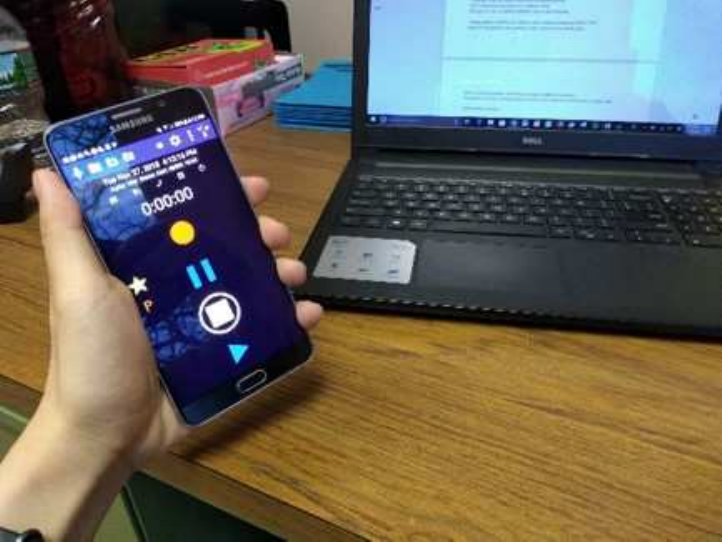}}%
	\hfill
	\subfigure[Idle on table]{%
		\label{fig:second}%
		\includegraphics[width=0.33\linewidth]{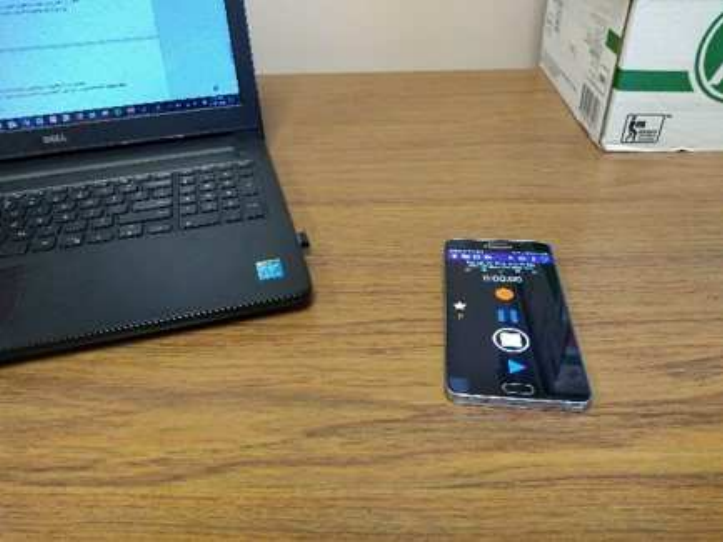}}%
	\hfill
	\subfigure[Stored in pocket]{%
		\label{fig:third}%
		\includegraphics[width=0.33\linewidth]{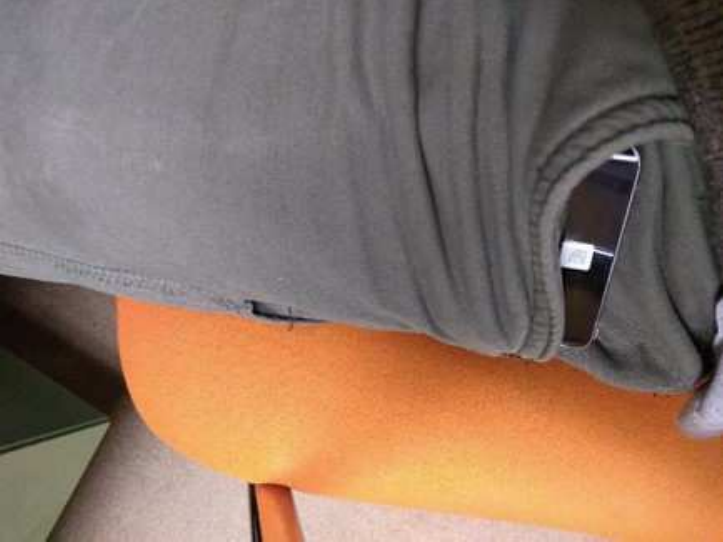}}%
	\vspace{-2mm}
	\caption{Testing conditions for preliminary environmental differentiation experiments.}
	\vspace{-5mm}
	\label{fig:test}
\end{figure}

\begin{equation}
K = \frac{-(p_1 - p_0)}{(V_1 - V_0)/V_0} = \frac{\Delta P}{\Delta V/V_0},
\end{equation}

\noindent
for a given differential change in pressure \(\Delta P\) and volume \(\Delta V\) relative to an initial volume \(V_0\). The introduction of a stable additive density \(V_1\) and fluctuating pressure increase \(p_1\) by the user changes K to a dynamic set of elasticity constants. This produces a range of acoustic patterns representative of how the device is held by a specific individual at the time of measurement, uniquely shaped by hand contour, posture, grip pressure, and behavior. Note that this model is an incomplete explanation as it does not account for a distributed application of pressure from different focal points of a user's hand. However, based on our observations, this explanation is sufficient for the purposes of conveying the intuition behind \textit{EchoLock}'s functionality.

\begin{figure}[t]
	\centering
	\includegraphics[width=2.1in]{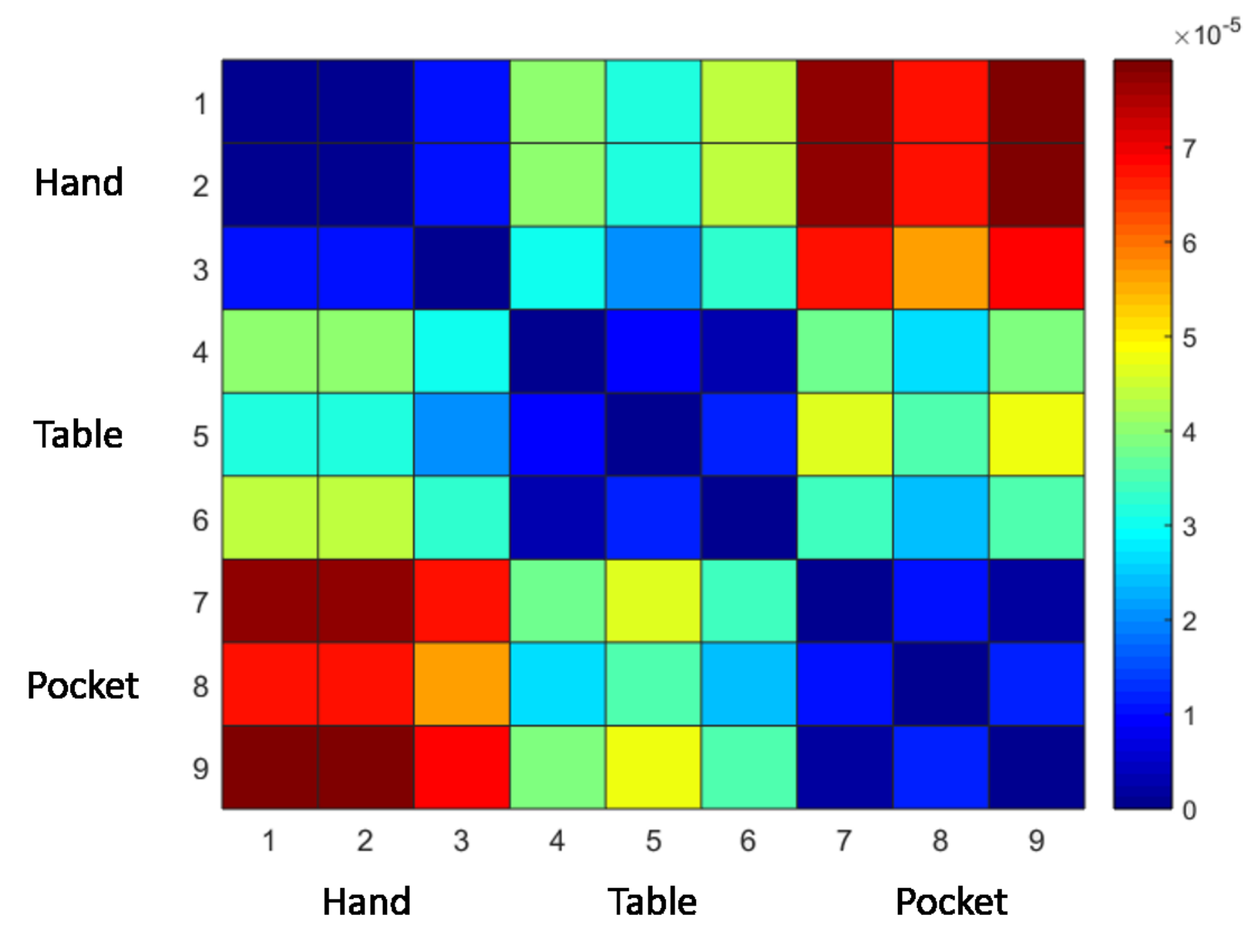}
	\vspace{-3mm}
	\caption{Euclidean distance of signal amplitudes recorded from different use case scenarios. Lower distances imply signal similarities.}
	\label{fig:study} 
\end{figure}

\begin{figure*}[t]
	\centering
	\includegraphics[width= 5.5in]{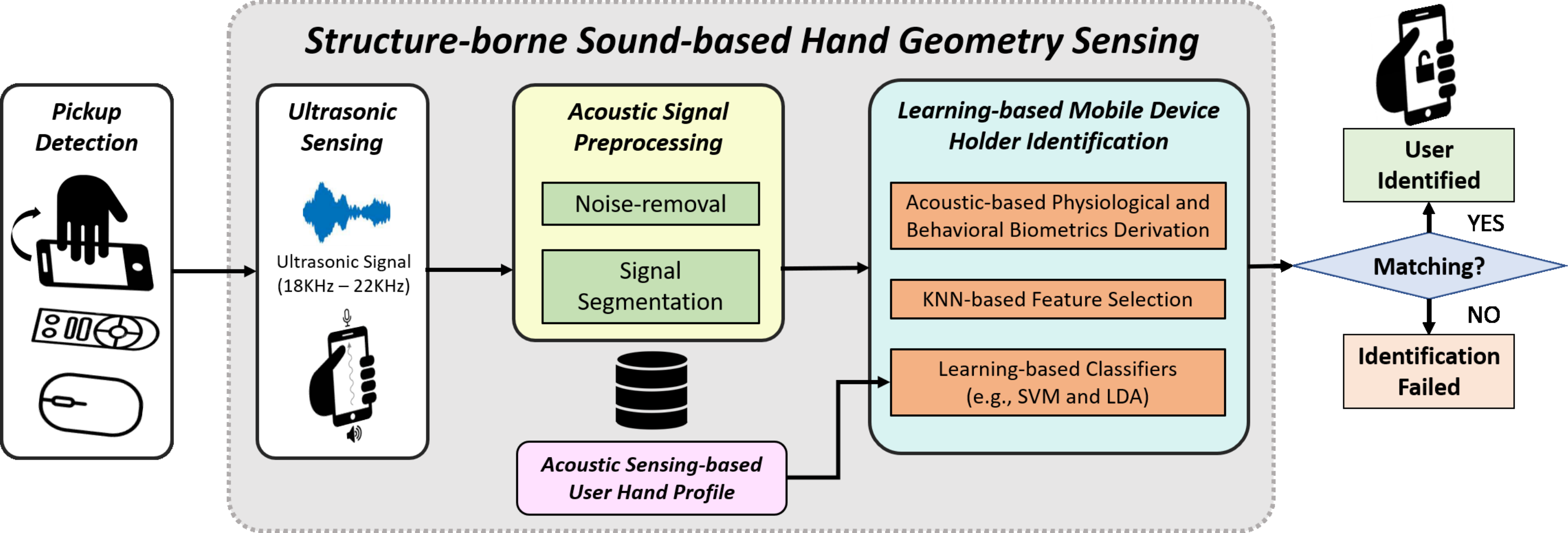}
	\caption{System Overview of EchoLock.} 
	\label{fig:system}
\end{figure*}
\subsection{Structure-borne Propagation Feasibility}
To establish the capabilities of structure-borne sound propagation, a simple preliminary experiment was performed to gauge the ability for a mobile device to ascertain environmental conditions using acoustic sensing. The observations from this experiment shaped the design of \textit{EchoLock} and provided insight into implementation feasibility and potential design challenges. Three different use cases were selected with the intent of demonstrating that conditions with distinct forces exerted upon the mobile device could be easily recognized. Figure \ref{fig:test} shows the particular experimental setup for each the three scenarios; having the mobile device resting in the user's hand, on a table, or within the user's pocket.

A Samsung Galaxy Note 5 was used to conduct the experiment. Due to its relatively large size, the compressing forces acted upon it are more pronounced as the sound propagation is prolonged by traveling through a larger medium. An inaudible chirp signal sweeping from 16 kHz to 22 kHz was emitted from the device speakers to induce vibration, which is then recorded by a single microphone near the top of the device. This captured recording is then examined for features that may identify environmental origins. The results of these experiments can be seen in Figure \ref{fig:study}. Three samples from each use case in Figure \ref{fig:test} were obtained and the average amplitude for each recording was extracted. The absolute difference was then computed for each combination of samples to measure statistical distinction between each use case.

Our findings show a clear relation between samples originating from different use cases. Samples native to the same use case displayed naturally low amplitude difference, which increased when compared to samples from foreign cases. Note across the diagonal that the difference is zero as these are comparisons between a sample with itself. This suggests that, even with minimal sensors and signal processing, structural-borne sound propagation is capable of communicating critical information about the immediate surroundings.

\vspace{-2mm}
\section{System Overview}
\textit{EchoLock} utilizes the properties of sound wave propagation to identify mobile device users, which can be applied to applications such as accessing banking accounts, managing notifications, and unlocking devices. In this section, we introduce the application of acoustic sensing to deduce the identity of the person holding the mobile device via structure-borne sound propagation, the attack model, and system overview.

\subsection{Attack Model}
\label{subsec:attack}

\textbf{Impersonation Attack.} The attacker attempts to mimic the holding posture of the legitimate user to gain access to the device. For impersonation attacks, the attacker may either be informed or uninformed. In the uninformed case, the attacker possesses no knowledge on how to circumvent the identification process and naively attempts to mimic the legitimate user's holding behavior. In the informed case, however, the attacker is explicitly aware of the legitimate user's authentication credentials in some form. This may be through passive observations of the user's hands or interactions such as handshaking. Physically faking the legitimate user's profile, however, requires applying forces to the device such that they create structural deformations similar to how the user's own hands would.

\textbf{Eavesdropping and Replay Attack.} The attacker attempts to steal acoustic credentials of the legitimate user by eavesdropping instances of identification attempts. This may be done by positioning a microphone near the user as \textit{EchoLock} is deployed. After obtaining an audio sample of a signal used to authenticate the user, the attacker gains possession of the targeted device and replays the audio sample via an external speaker. During ultrasonic sensing, the mobile device will transmit and record our acoustic signal. Successful replay of an eavesdropped sample necessitates the attacker to first suppress or prevent this signal to avoid overlapping with their own transmission, which is a non-trivial challenge.

\textbf{Jamming Attack.} The attacker in this scenario is focused on deliberate sabotage of genuine authentication attempts. This may be carried out by playing loud noise or ultrasonic frequencies near the user to disrupt the profile estimation procedure. The attacker does not necessarily need to know the user's credentials to jam the system. We assume in our assessment that the attacker will utilize ultrasonic frequencies to decrease the chances of detection by the ordinary user.

\subsection{Challenges and Requirements}
Although the results from our feasibility study are encouraging, they also highlight critical design challenges.
First, using a single built-in speaker and a microphone available on a mobile device to sense the user's complicated physical hand traits (e.g., hand geometry, finger placements and holding strength) is an unexplored area. 
Second, because the acoustic signals travel too fast compared to the small dimensions (i.e., sensing area) of mobile devices (e.g., 15 cm between a built-in speaker and microphone for a smartphone), the existing built-in microphone can only receive limited acoustic samples (< 20 samples~\cite{sun2018vskin,tung2016forcephone}) to describe a complete propagation. Using such limited acoustic information to describe the complicated interferences caused by people's hand is difficult. Third, the acoustic signals arriving at the microphone are the combination of structure-borne propagation and airborne propagation; how to separate or leverage these propagation signals to extract people's hand biometrics is challenging. Fourth, the environmental reflections of the acoustic sensing signals and the ambient noises corrupt the received sound and make the acoustic analysis of the user's holding hand even harder. Besides addressing these challenges, we also need to consider both security and usability when designing the system. In particular, the passive user input to our system should be hard to observe and imitate to meet security requirements. Moreover, our system needs to have a short acoustic sensing period to avoid causing additional delay and support the user's both left hand and right hand to maintain high usability.  


\subsection{System Architecture}
We designed \textit{EchoLock} as a fast and low effort user authentication scheme using existing hardware components to detect and process structure-borne sound waves. In order to authenticate a user's identity from these sound waves, we perform an acoustic signal preprocessing phase and a learning-based mobile device holder identification process as illustrated in Figure \ref{fig:system}.
\textit{EchoLock} uses the \textit{Pickup Detection} to continuously monitor an initiating action, such as the user picking up their device from a table or pocket, with the assumed intent that the user wishes to unlock the device.
Upon detection of this action, our system performs  \textit{Ultrasonic Sensing}, where the device's speaker briefly emits an inaudible acoustic signal (i.e., a chirp signal ranging from $18$kHz to $22$kHz) and immediately after records the user's reflection via onboard microphones. This procedure concludes within milliseconds, depending on signal design parameters we discuss in Section \ref{subsec:sigdesign}.

Then, this response undergoes our \textit{Acoustic Signal Preprocessing} phase, where we apply a bandpass filter from $18$kHz to $22$kHz and perform cross-correlation to remove noise and ensure that our obtained signal consists of structure-borne sound. 
Following the preprocessing, the system analyzes the signal to capture the user's physiological and behavioral characteristics in the \textit{Acoustic-based Physiological and Behavioral Biometrics Derivation}.
A series of statistical and frequency based features, further elaborated in Section \ref{sec:acoustic}, are identified based on their sensitivity to the forces, palm sizes and finger displacements of the user's hand and selected by KNN-based Feature Selection. 

These features are used to develop our learning-based classifiers, which are responsible for processes we refer to as \textit{profiling} and \textit{verification}. All extracted features representative of a specific user are compiled into a cell-based data structure and saved to a database of \textit{Acoustic Sensing-based
User Hand Profile}. This database is then referenced for a profile match when performing the user authentication. However, even known users may be difficult to verify due to behavioral inconsistencies in how they hold their devices. To combat this, we employ the \textit{Learning-based Classifiers} Support Vector Machine (SVM) and Linear Discriminant Analysis (LDA) learning techniques to examine numerical distance discrepancies and deduce the most probable user profile match. Once complete, \textit{EchoLock} finally concludes the process with an identification decision, performing a desired function, such as unlocking the device or denying access, based on the computed similarities between the measured features obtained from ultrasonic sensing and database of learned profiles. Because of the rapid propagation of sound through a physical medium, the entire process is conducted within milliseconds without any conscious exertions from the user.
\section{Low Effort Identification}
\label{sec:acoustic}

\begin{figure}[t]
	\centering
	\includegraphics[width= 2.5in]{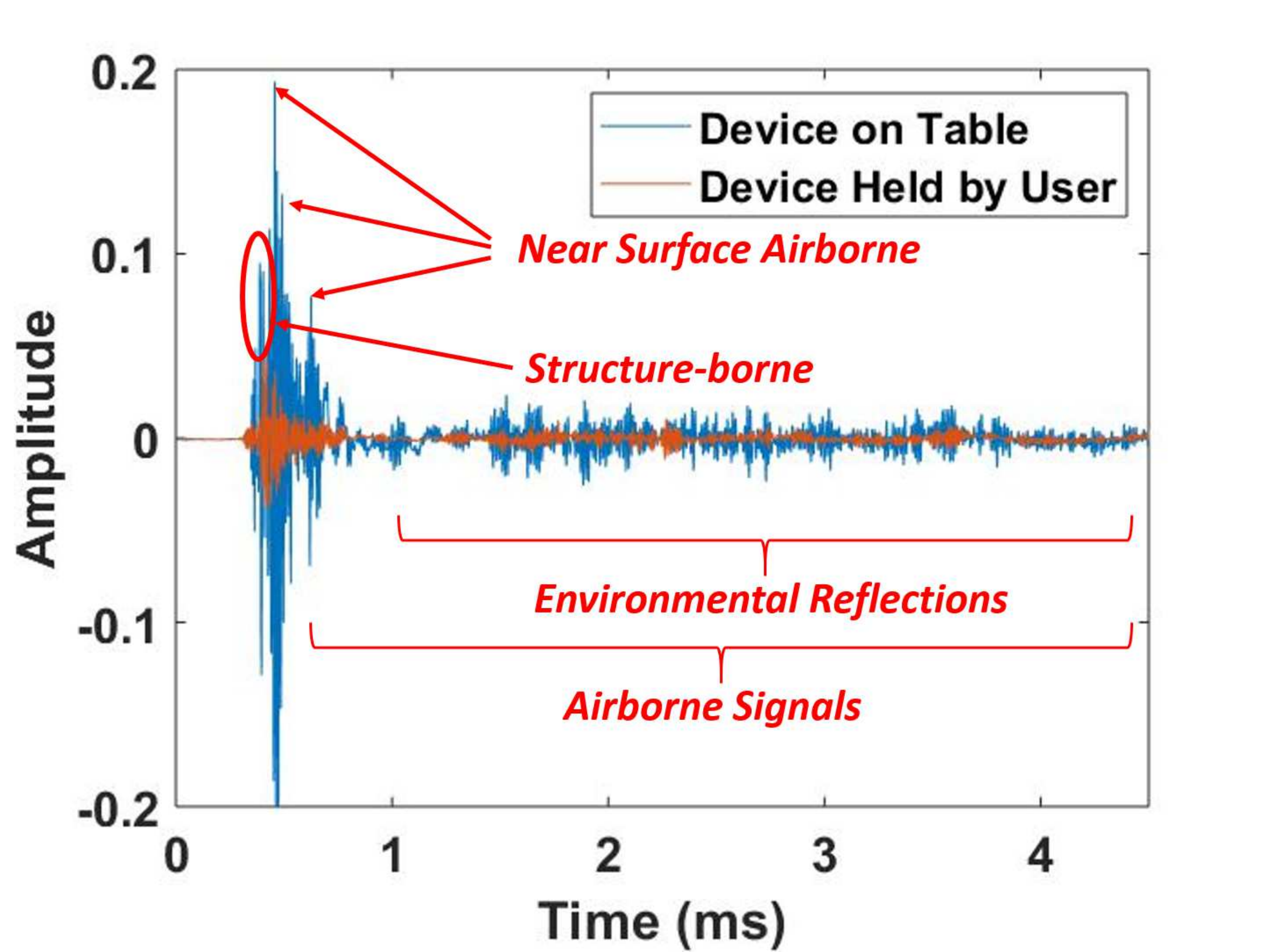}
	\vspace{-3mm}
	\caption{Structure-borne, near surface airborne sounds and environmental reflections.}
	\label{fig:structure_air}
	\vspace{-5mm}
\end{figure}

\subsection{Sound Propagation Separation}

When a user holds his or her mobile device, \textit{EchoLock} utilizes the device's speaker and microphone to send and receive acoustic sensing signals, which can capture the user's fine-grained hand biometrics. This is because the holding hand interacts (i.e. block and reflect) with the acoustic signals and creates unique patterns associated with the holding hand's physical characteristics, including the curvature and fat composition of the palm.

Intuitively, the structure-borne propagation of the acoustic signals is a good candidate for deriving the user's hand biometrics, since the structure-borne signals are directly affected by the contacting area and force of the hand holding the mobile device. However, extracting the structure-borne signals only is non-trivial because the built-in microphone captures both structure-borne and airborne signals. Existing studies propose to leverage the difference between the propagation speed of the structure-borne signal and that of the airborne signal to separate them~\cite{sun2018vskin,tung2016forcephone}.
Due to the small dimensions of mobile devices, the structure-borne signal only arrives at a very short time before the airborne signals (e.g., ~$0.20$ms).

In particular, the acoustic signals propagating along the paths close to surface of the mobile device may be blocked or reflected by the user's holding hand. 
Figure~\ref{fig:structure_air} shows the structure-borne and near surface airborne signals when a smartphone is placed on a table and held by a hand, respectively, when playing and recording a $1$ms short chirp signal using the built-in speaker and microphone. 

It is apparent that both structure-borne and near surface borne signals are affected significantly by a hand, however airborne sound is less reliable to estimate surrounding information due to distortions from the multipath effect. The minute delay of airborne sound may also introduce noise to subsequent structure-borne sound transmissions due to asynchronous arrival time, necessitating separation of the two. We note that while airborne sound is not directly used by our system to identify the user, it does provide an indirect masking effect against eavesdropping attempts by obscuring genuine structure-borne sound to external listeners.

\subsection{Acoustic Feature Extraction}
\label{subsec:feature_extraction}
\begin{figure}[t]
	\centering
	\includegraphics[width=2.2in]{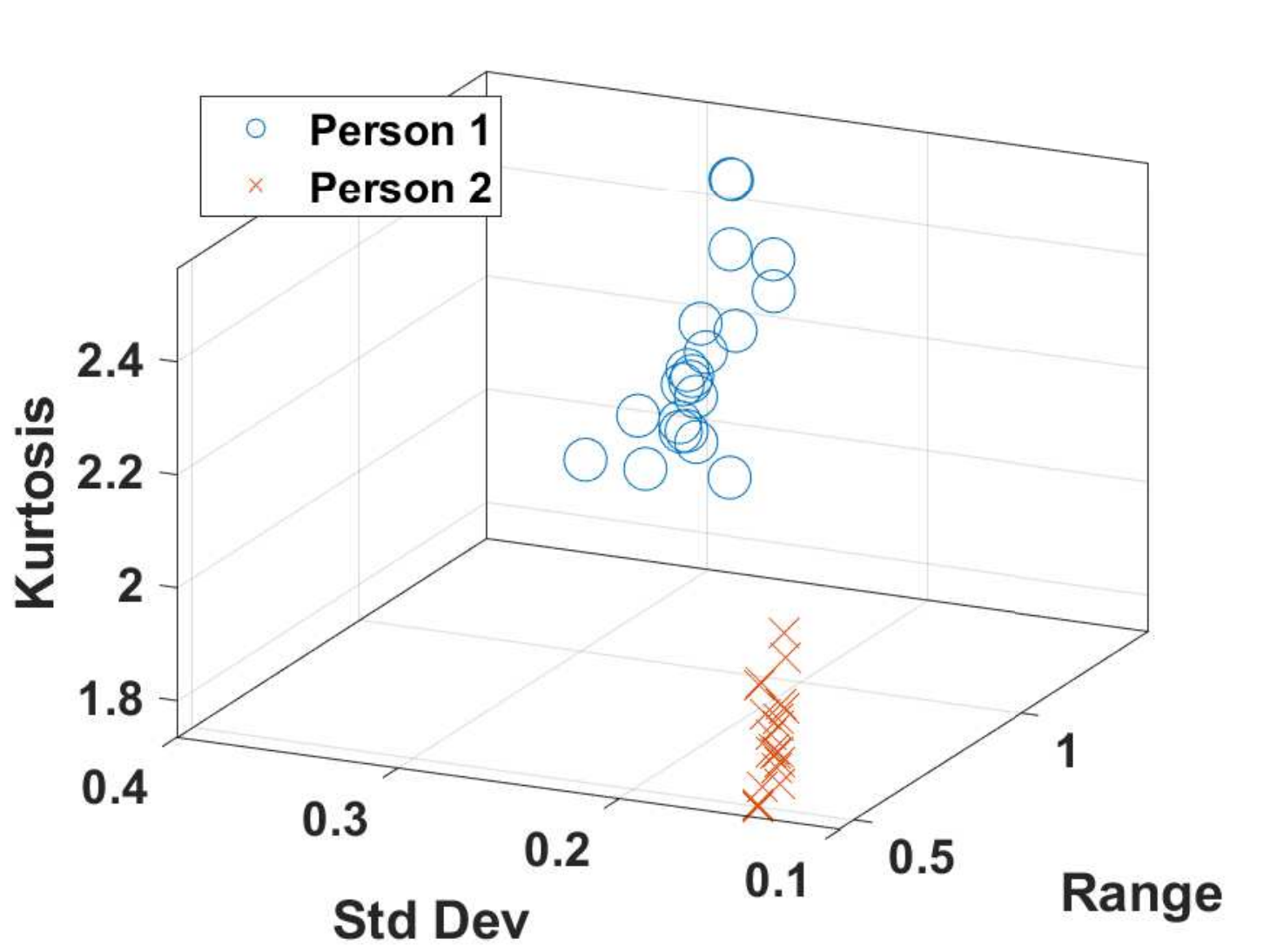}
	\vspace{-3mm}
	\caption{Illustration of the time-domain statistical features to differentiate two people's hands.}
	\label{fig:feature_ex}
	\vspace{-5mm}
\end{figure}

After receiving the designated signal, the system extracts from it unique features to analyze the interferences caused by the user's hand and derive a biometric profile, which integrates both physiological and behavioral traits.
A series of candidate features are identified for their potential responsiveness to different user hand biometrics. These features include statistical properties in the time domain, the spectral points in the frequency domain, and acoustic properties such as Mel-Frequency Cepstral Coefficient (MFCC) and Chromagram features.

\textbf{Time-domain Statistic Features.} In the time domain, we choose to analyze signals by its statistics including \textit{mean}, \textit{standard deviation}, \textit{maximum}, \textit{minimum}, \textit{range}, \textit{kurtosis} and \textit{skewness}. We also estimate the signal's distribution by calculating ~\textit{second quantile}, \textit{third quantile}, \textit{fourth quantile} and \textit{signal dispersion}. Additionally, we examine \textit{peak change} by deriving the index position of the data point that deviates most significantly from the statistical average. 
Figure~\ref{fig:feature_ex} shows an example of the features with the standard deviation, signal dispersion, and range of the received acoustic signal, illustrating the statistical features can effectively differentiate the user's holding hand.

\begin{figure}[t]
	\centering
	\includegraphics[width=2in]{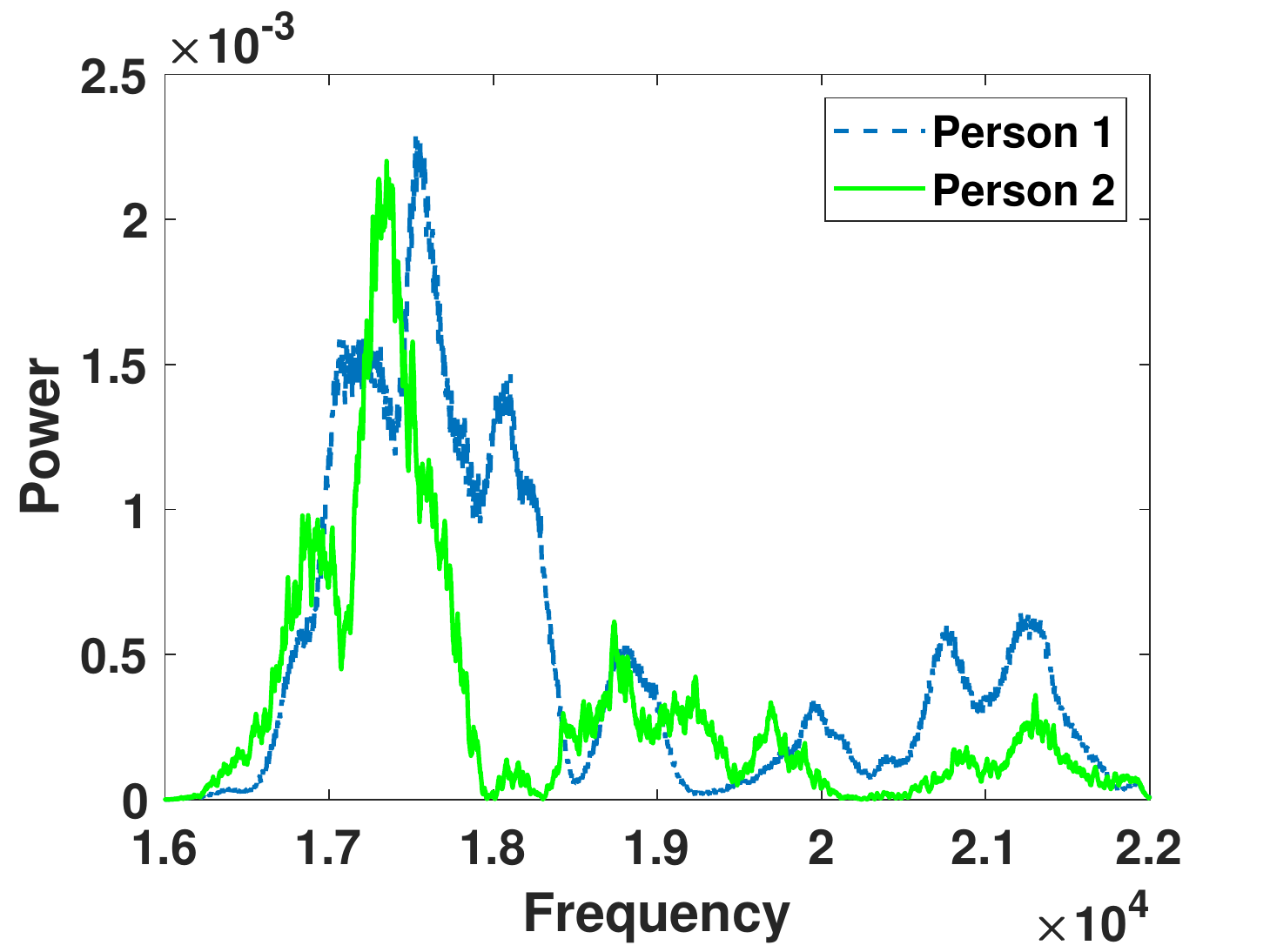}
	\vspace{-3mm}
	\caption{Spectral analysis of the received acoustic signal for two people's holding hands.}
	\vspace{-5mm}
	\label{fig:feature_ex_3}
\end{figure}

\textbf{Frequency-based Features.} In the frequency domain, we apply Fast Fourier Transformation (FFT) to the received acoustic signal and derive 256 spectral points to capture the unique characteristics of the user's holding hand in the frequency domain. This is because the holding hand can be considered as a filter, which results in suppressing some frequencies while not affecting others. Figure~\ref{fig:feature_ex_3} shows an example of the received sound in the frequency domain, where the frequencies between $19$kHz and $21$kHz are affected by the user significantly.

\textbf{Acoustic Features.} We also derive the acoustic features from the received sound using the MFCC~\cite{logan2000mel} and Chromagram~\cite{muller2005audio}. MFCC features are normally applied in speech processing studies to describe the short-term power spectrum of the speech sound and are good for reflecting both the linear and non-linear properties of the sound's dynamics. Chromagram, often referred to as "pitch class profiles", is traditionally utilized to analyze the harmonic and melodic characteristics of music and categorize the music pitches into twelve categories. We have observed the sensitivity of the MFCC and Chromagram to be sensitive enough to respond to physical biometrics as well. In this work, we derive $13$ MFCC features and $12$ chroma-based features to describe the different hand holding-related interferences to the sound.

\subsection{KNN-based Feature Selection}

Some features are more sensitive to the minute differences of different hands while the others may not be very effective at distinguishing between them. Moreover, mobile devices from different vendors may have their speaker and microphone embedded at different positions.
These hardware distinctions introduce further uncertainty when measuring the user using our features. In this work, we develop a K-nearest neighbour (KNN) based feature selection method find the more salient features for \textit{EchoLock} according to the mobile device when the user registers the system.

In particular, we apply KNN to each type of feature to obtain the clusters for different users. We then calculate the Euclidean distance of each feature point to its cluster centroid and that to centroids of other clusters. The purpose is to calculate the intra-cluster and inter-cluster distances to measure whether a feature is consistent for the same user and simultaneously distinct for different users. Next, we divide the average intra-cluster distance over the average inter-cluster distance and utilize an experimental threshold to select the features. The selected features based on KNN are not only sensitive to the user' hand holding activity but also resilient to the other factors such as acoustic noises.

\subsection{Learning-based Holder Verification}
\label{subsec:learning_verify}
We develop learning-based algorithms to learn the unique characteristics of the user's hand holding activity based on the derived acoustic sensing features and determine whether the current device holder is the legitimate user or not. In particular, we utilize Support Vector Machines (SVM) and Linear Discriminant Analysis (LDA) as two alternative classifiers to classify the mobile device holder a registered or unknown user. SVM relies on a hyperplane to divide the input acoustic sensing feature space into the categories with each representing a user. The hyperplane is determined during the training phase with the acoustic sensing data from the registered users. We use LIBSVM with a linear kernel to build the SVM classifier~\cite{chang2011libsvm}. LDA finds a linear combination of features that characterizes or discriminate the acoustic sensing data in different user classes and utilizes Bayes' theorem for classification. 


\begin{figure}[t]
	\centering
	\includegraphics[width=2in]{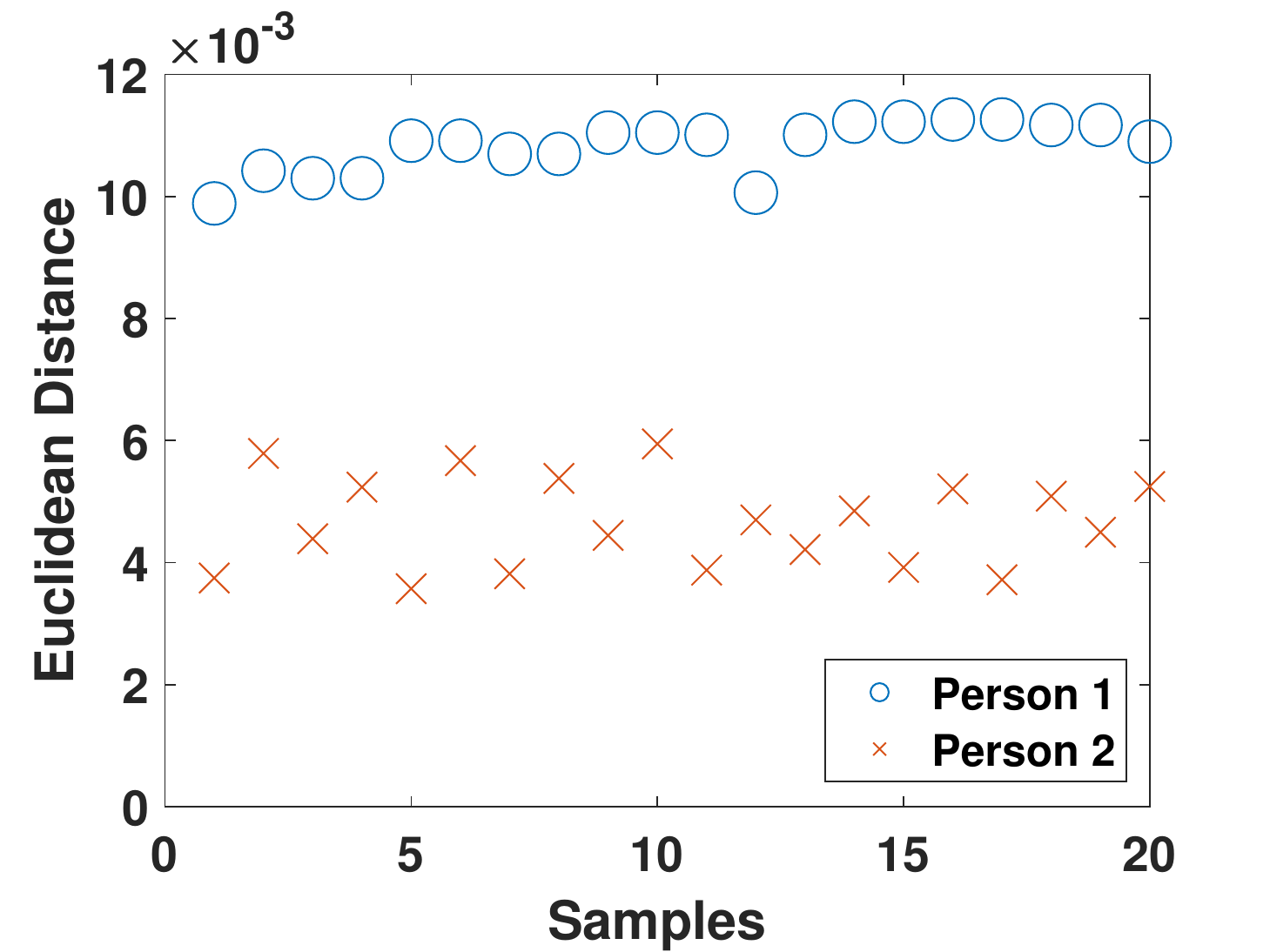}
	\vspace{-3mm}
	\caption{Euclidean distance of the standard deviation feature for two users relative to a separate instance of the feature for the second user.}
	\label{fig:feature_ex_2}
	\vspace{-5mm}
\end{figure}

The classifiers are trained during the user profile construction phase that is detailed in Section~\ref{subsec:user_profile}. During the user verification phase, \textit{Echolock} first classifies the testing data to one user based on the user profile. Then, our algorithm utilizes the prediction probabilities returned by the classifier as a confidence level and applies a threshold-based method to examine the classification results. If the confidence level of the classification is above the threshold, the user is identified. Otherwise, our system will determine it as ``unknown'' and respond accordingly. This can be used for multiple applications, such as immediately adjusting user settings when a registered user is detected, or locking devices when an unknown user attempts to gain access. 


	\section{Implementation}
\label{sec:implementation}

\subsection{Acoustic Sensing Signal Design}
\label{subsec:sigdesign}

The specific design of the acoustic signal emitted is paramount to the performance of \textit{EchoLock}. Consequentially, measuring biometric properties of a given person accurately using only sound propagation requires our signal to satisfy several design criteria:
\begin{itemize}[leftmargin=*]
	\item The signal must be designed in such a way to easily distinguish between the structure-borne and airborne sound propagation.
	
	\item The transmitted signal should be recognizable such that it can be easily identified and segmented amid interference from ambient noise and other acoustic disruptions. This is doubly advantageous for user identification as the procedure should be as fast as possible to avoid inconveniencing the user.
	
	\item The signal should fall within a safe frequency range inaudible to average human hearing. This is primarily for the purpose of usability as a noticeably audible signal may pose a nuisance to some users if they are subjected to it every time they wish to verify their identity. Generally speaking, $16$kHz is the upper bound of easily detectable sound for ordinary adults~\cite{ashihara2007hearingthresh}.
	
	\item The signal must be able to be deployed on a COTS device, limiting the viable transmission frequency range. Android devices, for example, are reported to have a maximum sampling rate of around $44$kHz, limiting a practical signal to $22$kHz at most~\cite{androiddev}. 
	Many devices, however, exhibit considerable attenuation problems when transmitting frequencies exceeding $20$kHz due to hardware imperfections in onboard speakers ~\cite{sun2018vskin, tung2015echotag, tung2016forcephone, zhou2018echoprint}. 
	However, \textit{EchoLock} actually leverages this property as a means of security as these hardware imperfections introduce variability that is consistent with the device origin, making it difficult for attacking signals to trigger false-positive results due to the challenges of mimicking degradation specific to a particular device.
\end{itemize}


With these considerations, we choose a chirp signal sweeping from 18kHz to 22kHz. This chirp consists of 1200 samples, equating to a 25ms duration at a 48kHz sampling frequency. During ultrasonic sensing, the recorded structure-borne propagation of the chirp signal will be imbedded with information on the user's hand geometry. While a shorter signal minimizes exposure to environmental reflections, it also limits the signal-to-noise ratio (SNR). To balance these two considerations, we transmit a series of consecutive chirps, separated by 25ms of empty buffers to stagger the arrival of airborne reflections from structure-borne sound. We refer to this design as a $n$-chirp sequence where $n$ is the number of repeating chirps in the signal used during implementation. By utilizing multiple chirps, we can also gather multiple user samples in a single ultrasonic sensing instance. This leads to a natural tradeoff dilemma between higher classification accuracy and shorter time delays. We show in further detail the performance accuracy for increasingly large $n$ values in Section \ref{subsec:impacts}.

\subsection{Noise Removal and Signal Segmentation}
%

\begin{figure}[t]
	\centering
	\subfigure[Full Recording]{%
		\label{fig:segment_before}%
		\includegraphics[width=0.49\linewidth]{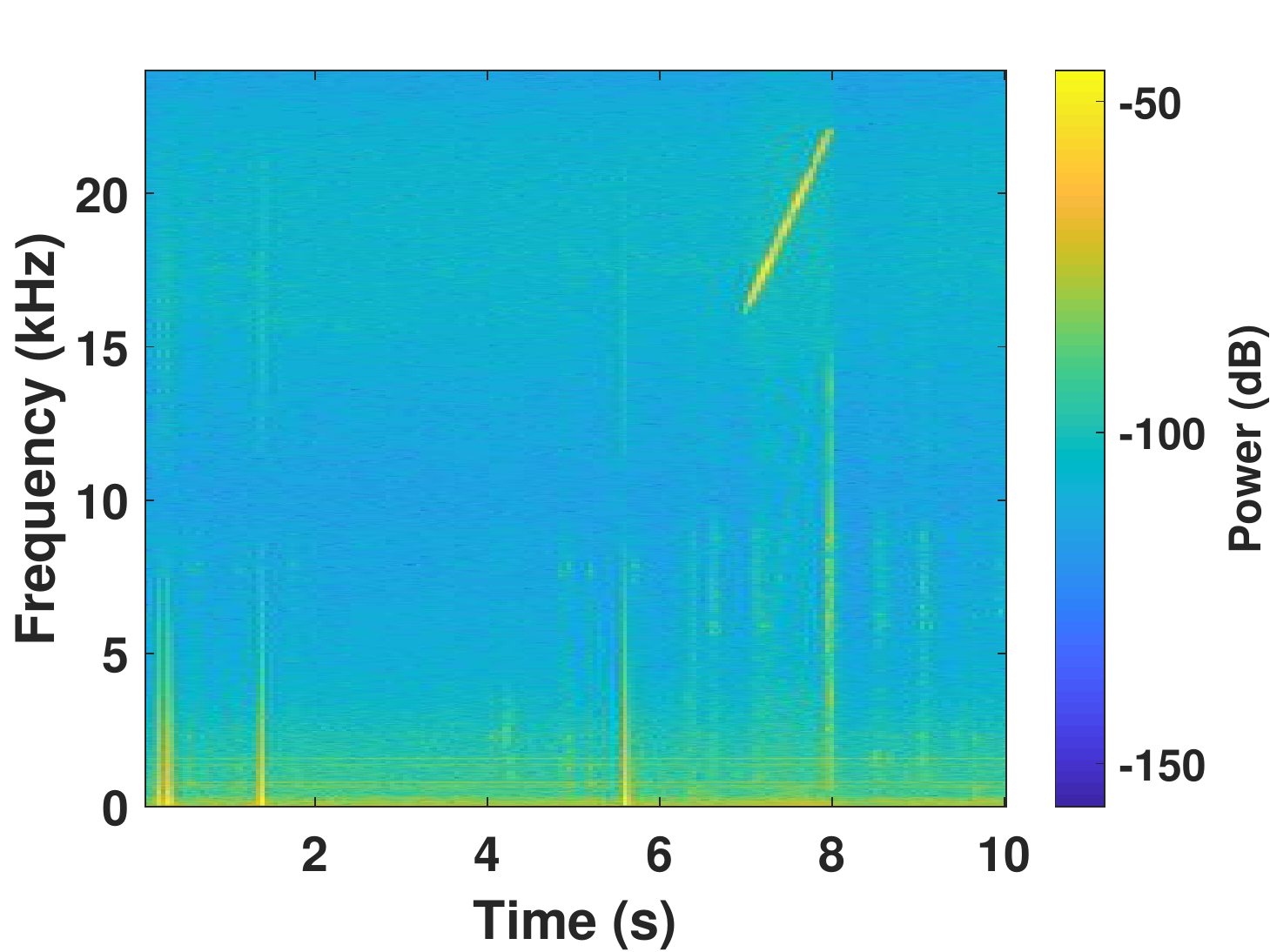}}%
	\hfill
	\subfigure[Extracted Signal]{%
		\label{fig:segment_after}%
		\includegraphics[width=0.49\linewidth]{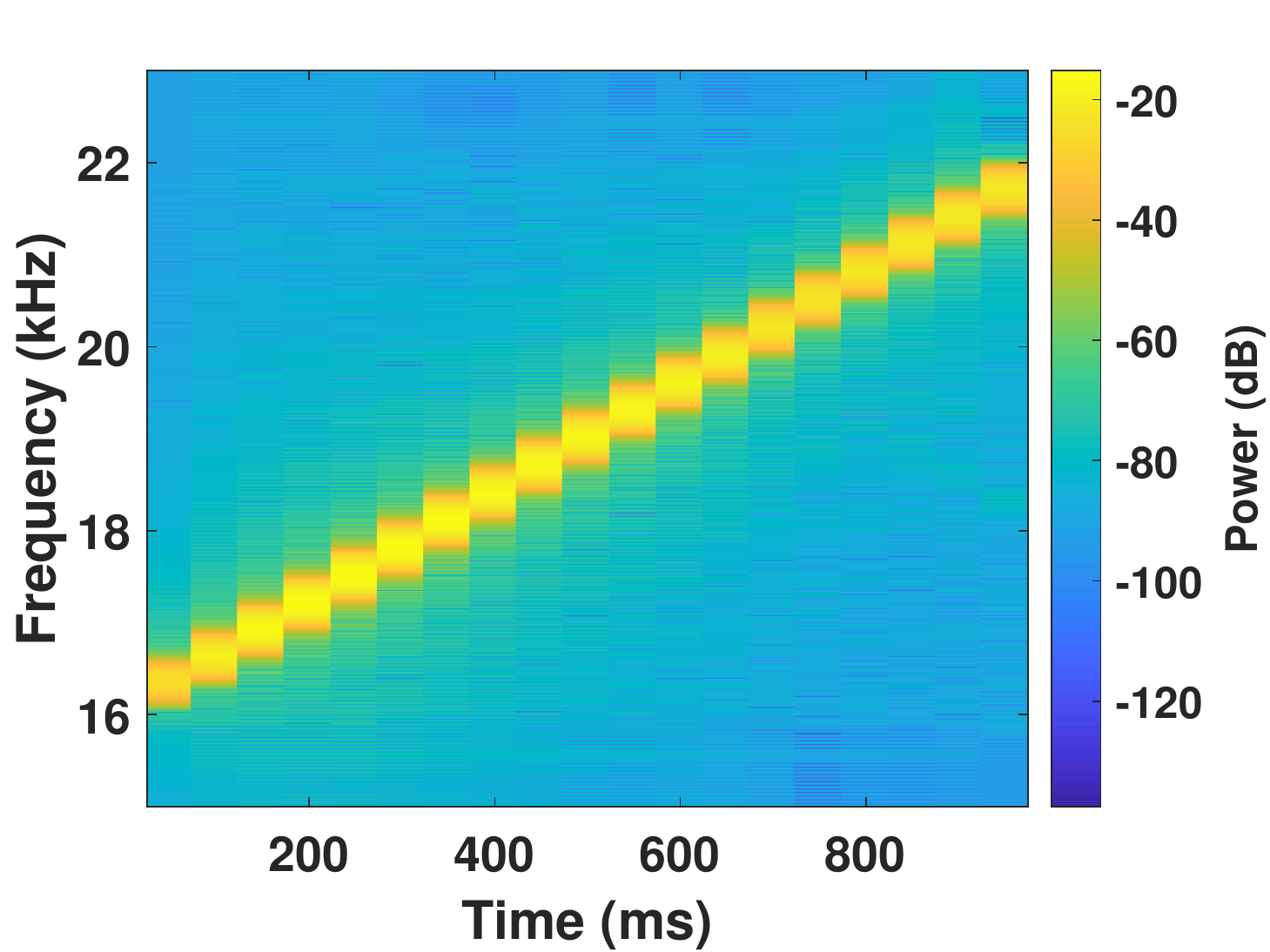}}%
	\vspace{-3mm}
	\caption{Signal segmentation on a microphone recording containing the desired acoustic features.}
	\vspace{-5mm}
	\label{fig:isolate}
\end{figure}

As described previously in Section 3, structure-borne sound propagation is capable of conveying information representative of the surrounding environment. In order to extract this information, it is necessary to remove the interference generated by various sources (i.e. irregular vibration patterns and ambient background noise) and determine the precise segments of the acoustic signal containing the target structural sound pattern.
A naive assertion could be made suggesting that the system can be designed to transmit and recapture the sound in a synchronized time interval based on the speed of sound through a physical medium. Not only does this strategy disregard the minute, unpredictable time delays introduced as a result of multiple layers of hardware and software operations, it also requires precise knowledge regarding the positioning of speakers and microphones relative to each other, as well as the particular material the device is constructed from. These conditions are neither consistent nor easy to acquire across multiple devices, thus a more reliable procedure must be developed. 

\textbf{Noise Removal.} Thanks to the design of our acoustic sensing signal, we can separate our transmission from interference by using a band-pass filter.
In particular, we design a band-pass filter with the pass band through $18$kHz to $22$kHz, which is the expected frequency range of the transmitted chirp signal. We apply a Butterworth low-pass and high-pass filter at the specified frequencies to achieve this.

\textbf{Signal Segmentation.}
After noise filtering, we need to identify the precise segment containing the structure-borne sound.
This can be accomplished by using cross-correlation to locate similarities between the recorded acoustic signal with the original transmission. 
Figure \ref{fig:isolate} shows the acoustic signal before and after the proposed noise removal and signal segmentation,
using a volunteer holding a Samsung Galaxy Note 5 as an example. The figure demonstrates that the acoustic signal received is filtered and and truncated to contain only 
the desired frequency range of the signal transmitted, suggesting that our methods can successfully remove noises and precisely extract the information containing the user's hand characteristics.


\subsection{User Profile Construction}
\label{subsec:user_profile}
EchoLock requires the user to provide the training data for user profile construction during registration. 
The acoustic sensing signals are emitted during this process, and the acoustic sensing features (in Section~\ref{subsec:feature_extraction}) are derived and labeled to train the learning-based classifier as described in Section \ref{subsec:learning_verify}.
An existing data set of anonymized user profiles is included to serve as negative labels when classifying the user during identification attempts. Registering multiple users to the same device also serves to expand this data set.
To ensure robustness and low false negative rates, the user is advised to vary holding behavior multiple times rather than remain still to train the data. This can be verified using motion sensors to detect change in device holding posture. 

\section{Performance Evaluation}
\subsection{Experimental Setup} 
We study the performance of \textit{EchoLock} in a variety of common use case scenarios as well as on several mobile devices. Our experiments test the capability of our system to lock or unlock access to mobile devices as an example application, approved by our institute IRB. We present our findings and detailed analysis below.

\begin{figure}[h]
	\centering
	\subfigure[Nexus 5]{%
		\label{fig:nexus}%
		\includegraphics[width=0.32\linewidth]{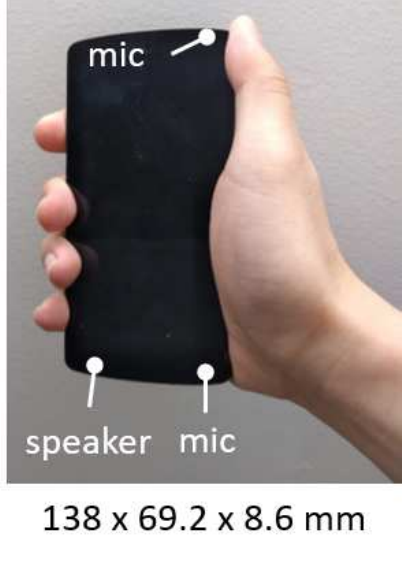}}%
	\hfill
	\subfigure[Galaxy Note 5]{%
		\label{fig:galaxy}%
		\includegraphics[width=0.32\linewidth]{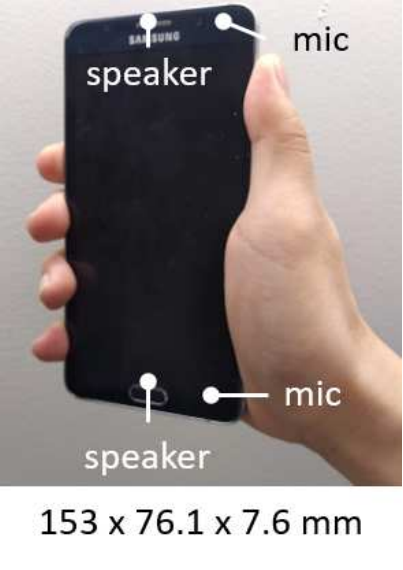}}%
	\hfill
	\subfigure[Galaxy Tab A]{%
		\label{fig:tab}%
		\includegraphics[width=0.32\linewidth]{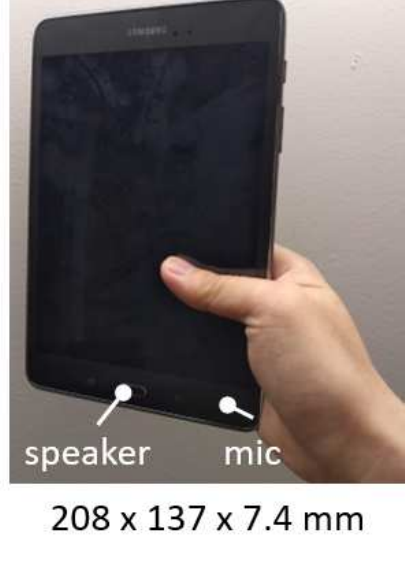}}%
	\hfill
	\caption{Mobile device specifications and locations of key sensor components.}
	\vspace{-3mm}
	\label{fig:devices}
\end{figure}

\textbf{Experimental Setup.} 
\label{subsec:exp}
A prototype application for \textit{EchoLock} was developed for use on Android devices. 
Three Android devices, the Google Nexus 5, Galaxy Note 5, and Galaxy Tab A, were selected for their varied designs (e.g., speaker and microphone positions) and size discrepancies, pictured in Figure \ref{fig:devices}. Both the Nexus 5 and Galaxy Note 5 devices include two onboard microphones whereas the Galaxy Tab A is equipped with one. 
Two types of environments explored were office and public settings. Office settings consist of quiet, enclosed spaces with minimal disturbances whereas public settings are locations with large volumes of people and traffic. 
We maintain an average noise level of approximately $30$dB and $60$dB for the two environments, respectively. Sources of noise for the public environment include nearby conversations, walking, and dining. We gauge the ability of our system to accurately identify the user in the face of these obstacles. 
Additionally, we also investigate the impact of accessories that may transform the properties of the user's hand or device structure, such as gloves or smartphone cases (examples in Appendix). From these identified factors, we devised several scenarios to study.


\textbf{Data Collection.} 
8 use case experiments were devised in total, divided into 3 general categories. The first and second categories study performance differences across device models and surrounding noise levels, respectively, comprised of the Cartesian product of our 3 devices and 2 environmental conditions. The third category considers usage via indirect physical contact, which includes when the user has equipped a protective case to their device and when the user is wearing a glove while holding their device. To reduce the number of factors at play during our third category of tests, we confine our study to the Galaxy Note 5 device in office settings. 
No specific instruction was provided to the participants on how to hold the device to encourage more natural interactions. However, we find that almost all participants favored holding postures similar to those shown in Figure \ref{fig:devices}. This is both beneficial and challenging as this adds consistency to our data set while also making it less simple to differentiate usage behaviors.
We recruit 20 volunteers, 14 males and 6 females ranging from ages 18-35, to participate in our study. We collect 40 $n$-chirp sequences, where $n$=10, for each test case based on the procedure in Section~\ref{subsec:user_profile} for a total of 64,000 hand geometry samples. The profiles of all volunteers collectively act as the negative label during classification, with the exception of the target user undergoing identification. Our data collection process spanned a 2 month period with volunteers providing data across different days. While this data is used for user identification purposes, we do not consider it to be personally identifiable information (as mentioned in Table~\ref{tab:quality}). Nonetheless, we are currently maintaining this data privately and do not plan to release it publicly as a precaution.

\subsection{Evaluation Metrics}
We describe the accuracy of our system by evaluating two criteria from our results, \textit{precision} and \textit{recall}. Precision is defined as the percentage of True Positive (TP) classifications out of all positive classifications recorded, notated as $P = \frac{TP}{TP+FP}$ where FP is the false positive rate. Recall is defined as $R = \frac{TP}{TP+FN}$ or the percentage of true positive classifications out of all target class instances. For our purposes, higher precision describes lower probability for different people to be mistaken for the legitimate user while higher recall describes the lower probability that the legitimate user is misidentified as someone else.

\begin{figure}[t]
	\centering
	\includegraphics[width=\linewidth]{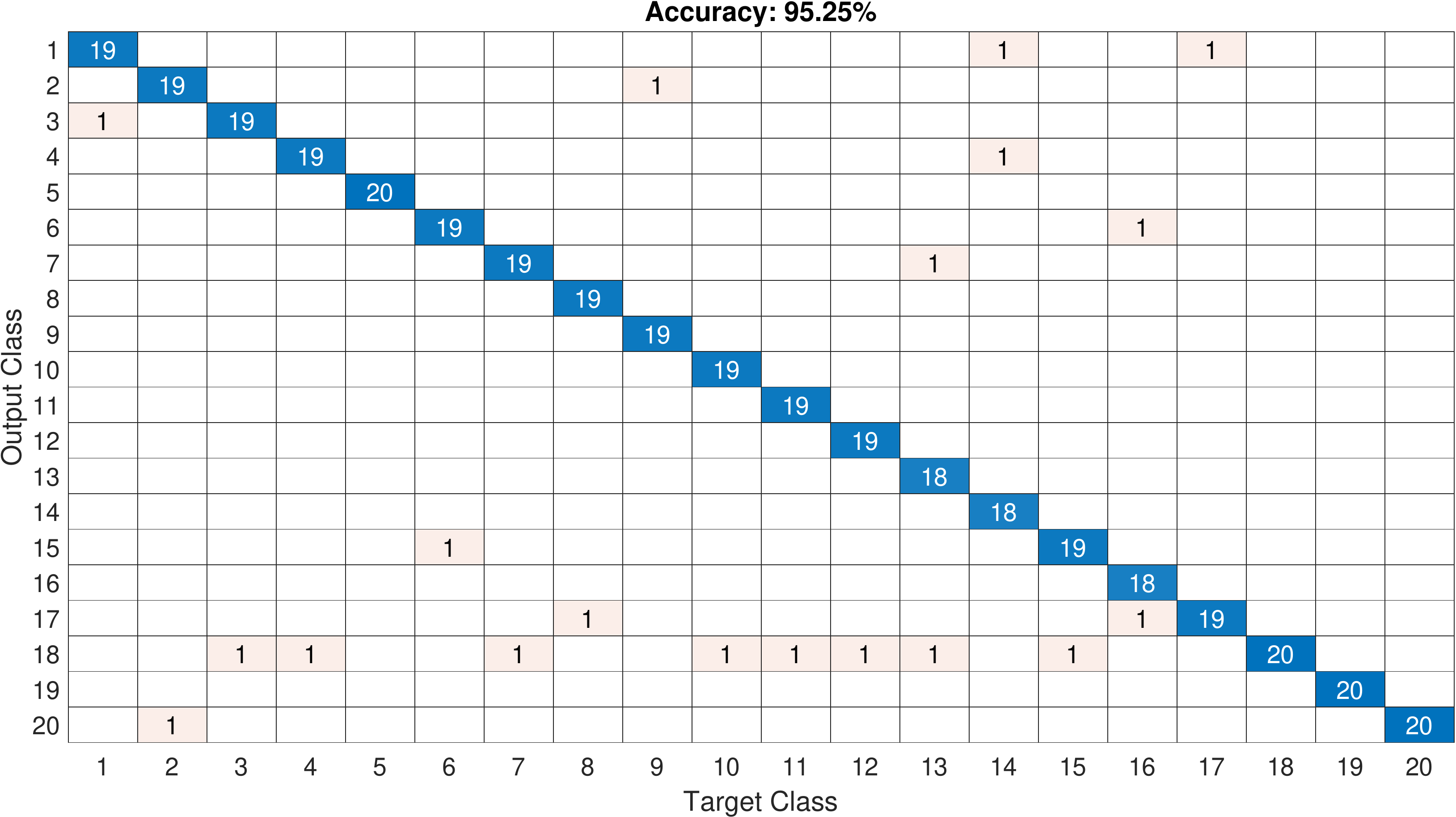}
	\caption{Confusion matrix of user identification when deploying \textit{EchoLock} on the Nexus 5.} 
	\vspace{-3mm}
	\label{fig:conf}
\end{figure}

\subsection{User Identification Performance}
From our signal processing procedures, we obtain several features used to identify the user and present them to a machine learning classifier for authentication. To evaluate the performance of our implementation, we consider Decision Trees (DT), Linear Discriminant Analysis (LDA), K-Nearest Neighbor (KNN), and Support Vector Machines (SVM) as our candidate classifiers. Figure \ref{fig:conf} shows an example confusion matrix demonstrating user identification accuracy for a group of 20 users in office settings. From our initial comparisons of each classifier's ability to distinguish between 5 different users, we observe LDA and SVM to demonstrate reasonable validation, showing accuracy of 95\% and 91\%, respectively, and smaller range compared to DT and KNN. As a result, we present our findings for these two classifiers for our extended evaluations. We choose 10-fold cross-validation during the training process to best utilize our data set and minimize selective bias, allocating 50\% for training and the remaining 50\% for testing.


\subsection{Impact Factor Study}
\label{subsec:impacts}

\begin{figure}[t]
	\centering
	
	\hfill
	\subfigure[SVM]{%
		\label{fig:performance_device_svm}%
		\includegraphics[width=0.49\linewidth]{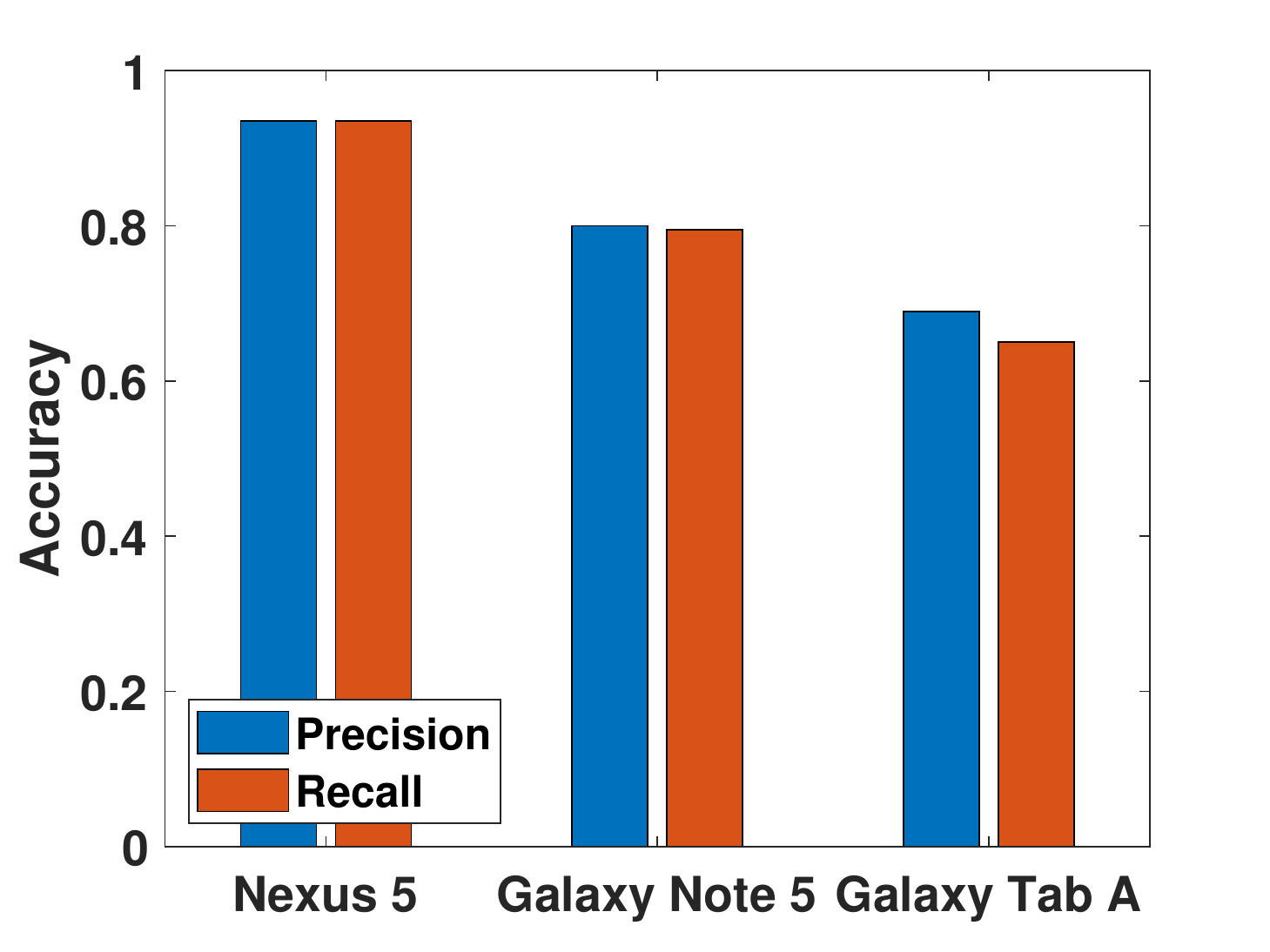}}%
	\hfill
	\subfigure[LDA]{%
		\label{fig:performance_device_lda}%
		\includegraphics[width=0.49\linewidth]{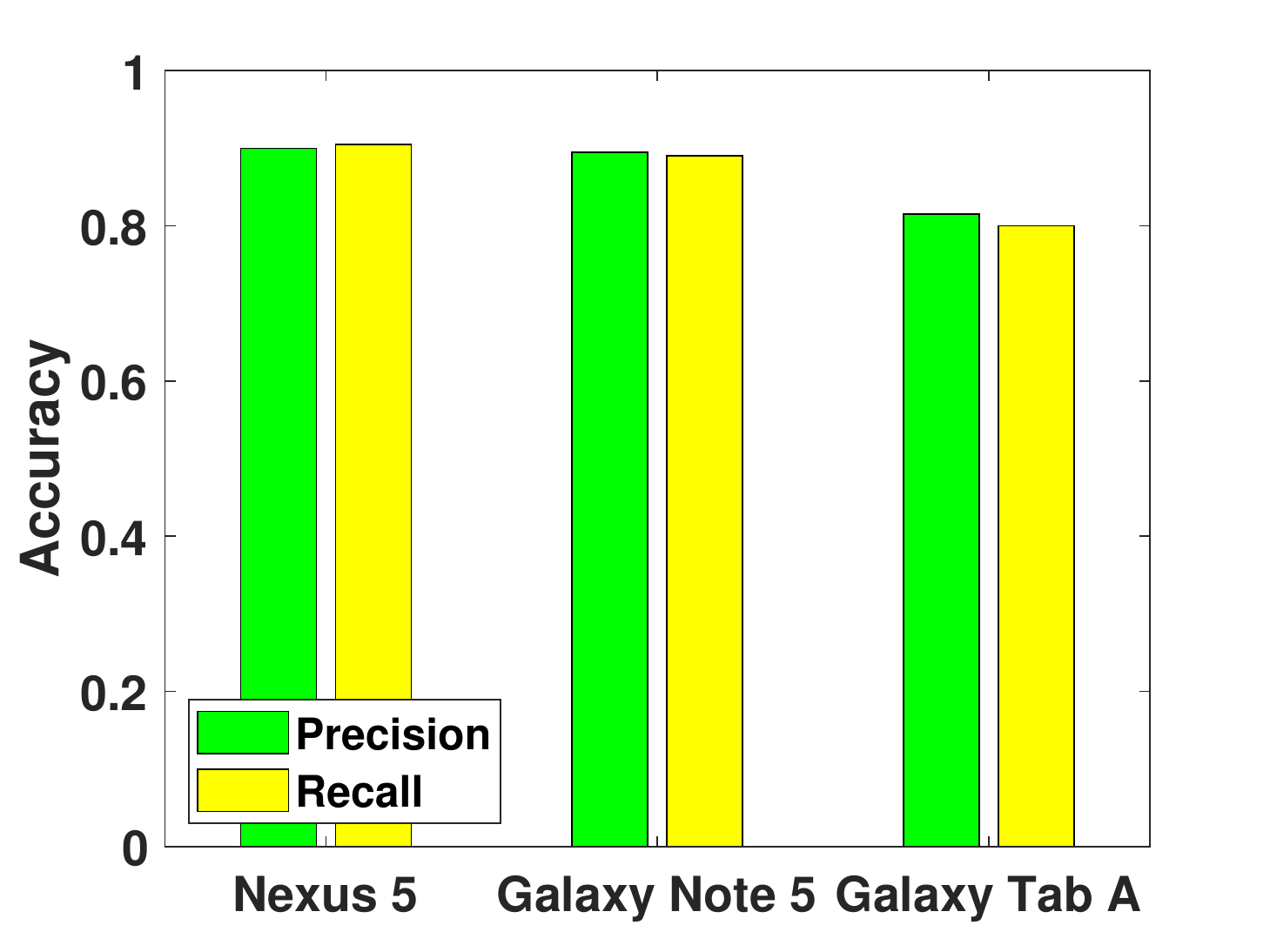}}%
	\hfill
	\vspace{-3mm}
	\caption{Performance for different mobile device models when used in office and public settings.}
	\vspace{-4mm}
	\label{fig:performance_device}
\end{figure}

\begin{figure}[t]
	\centering
	
	\hfill
	\subfigure[SVM]{%
		\label{fig:performance_enviro_svm}%
		\includegraphics[width=0.49\linewidth]{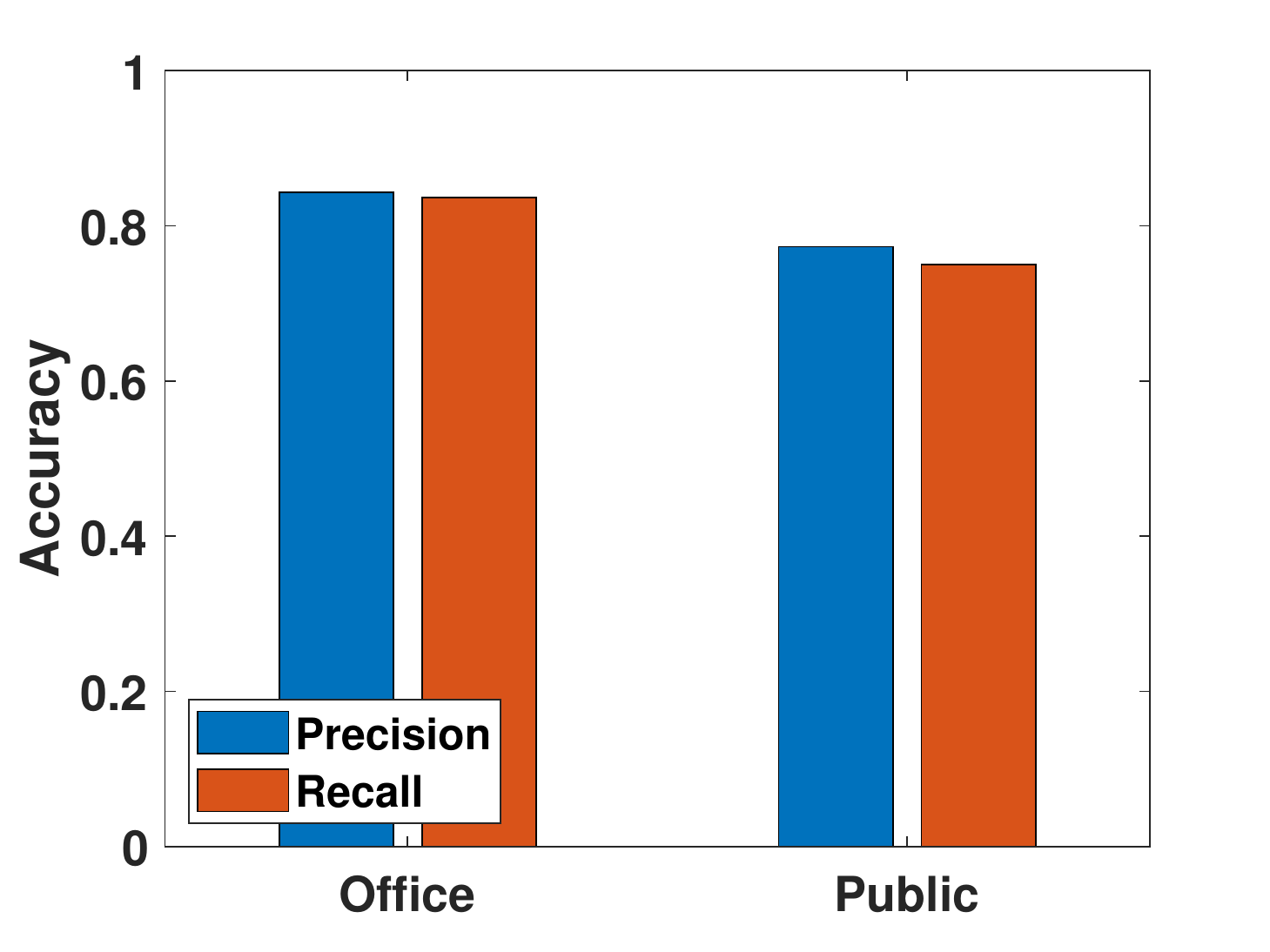}}%
	\hfill
	\subfigure[LDA]{%
		\label{fig:performance_enviro_lda}%
		\includegraphics[width=0.49\linewidth]{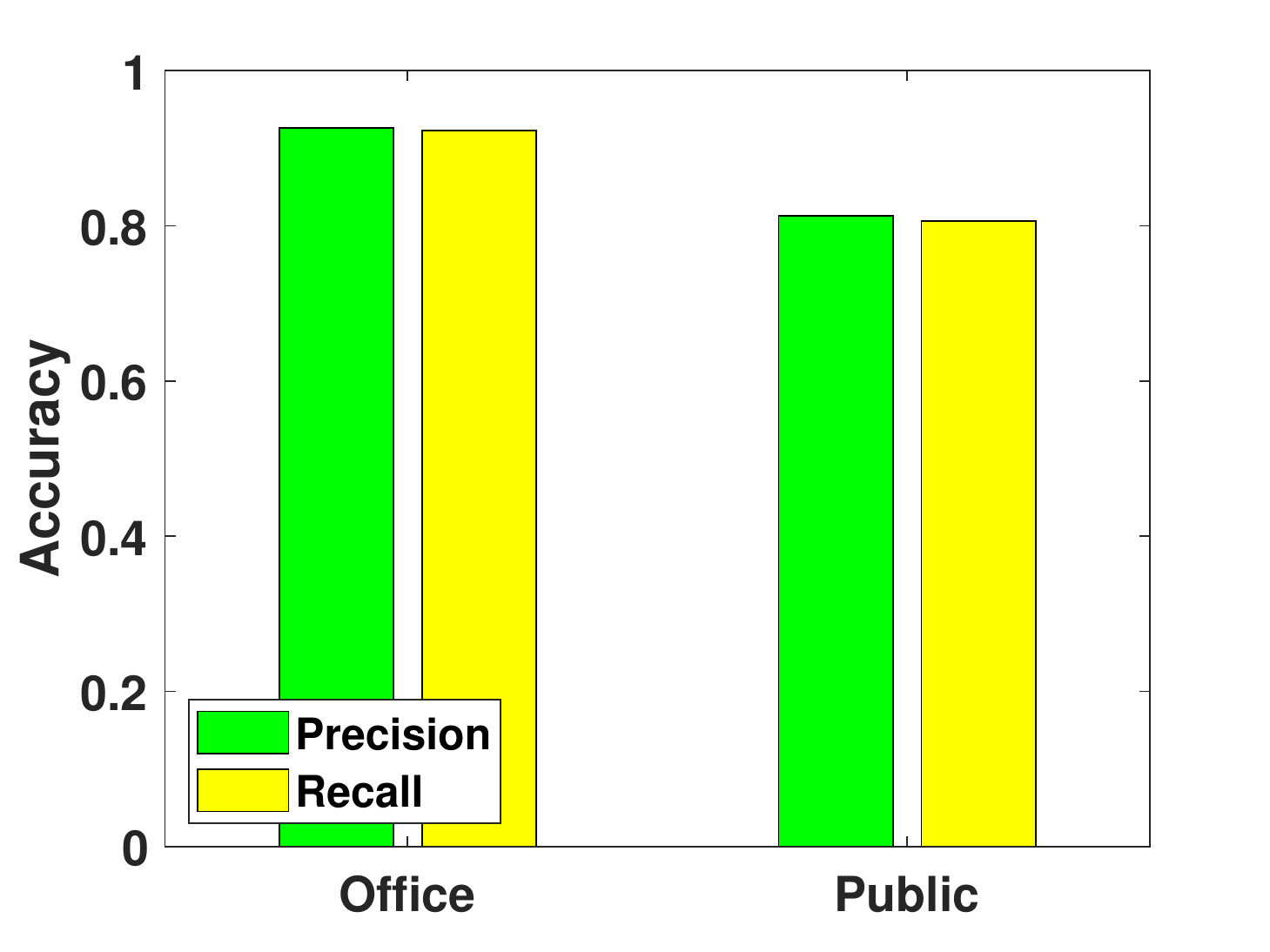}}%
	\hfill
	\vspace{-3mm}
	\caption{Performance when identifying users in different environments.}
	\vspace{-4mm}
	\label{fig:performance_enviro}
\end{figure}

\begin{figure}[t]
	\centering
	
	\hfill
	\subfigure[SVM]{%
		\label{fig:performance_indirect_svm}%
		\includegraphics[width=0.49\linewidth]{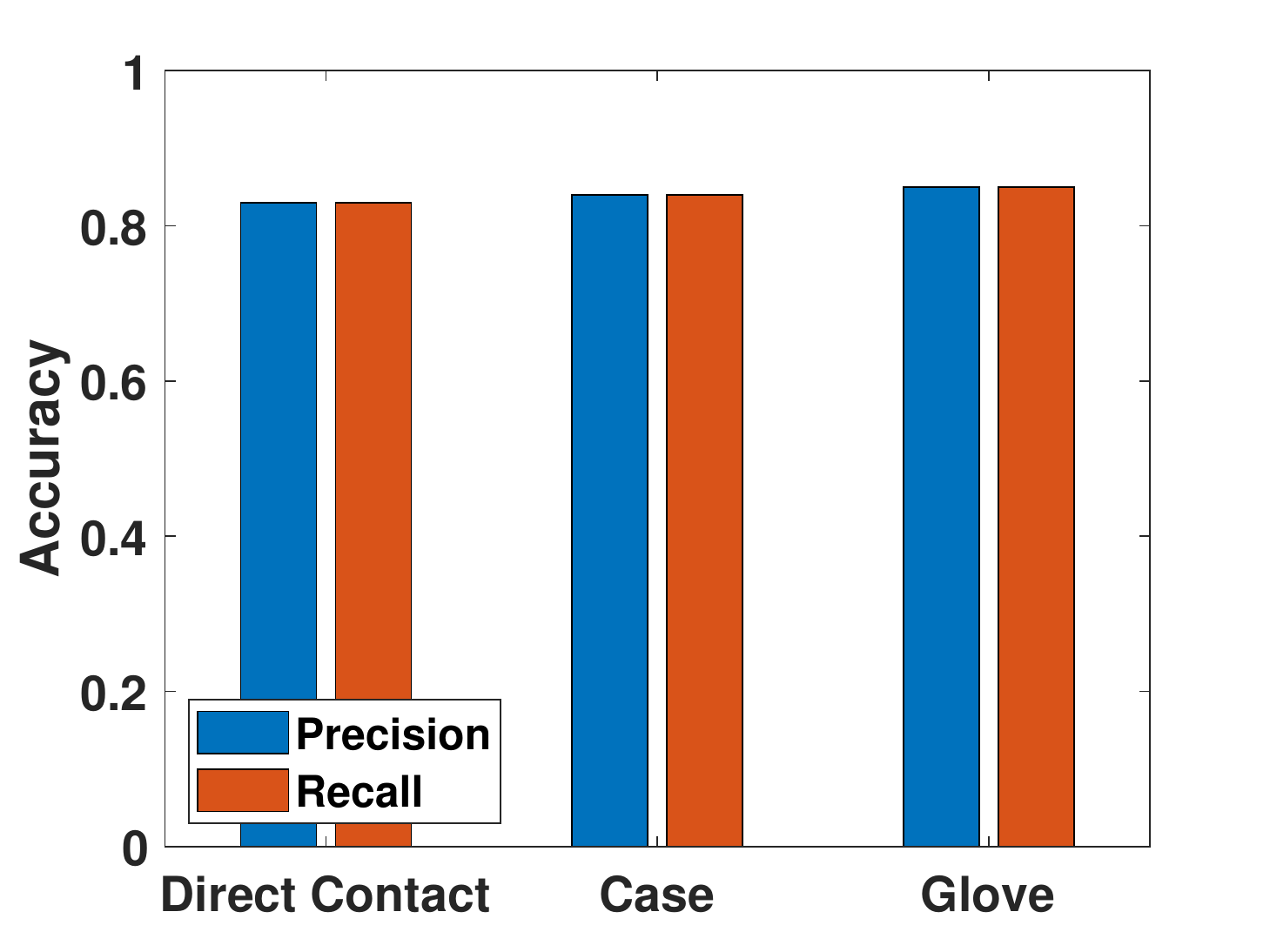}}%
	\hfill
	\subfigure[LDA]{%
		\label{fig:performance_indirect_lda}%
		\includegraphics[width=0.49\linewidth]{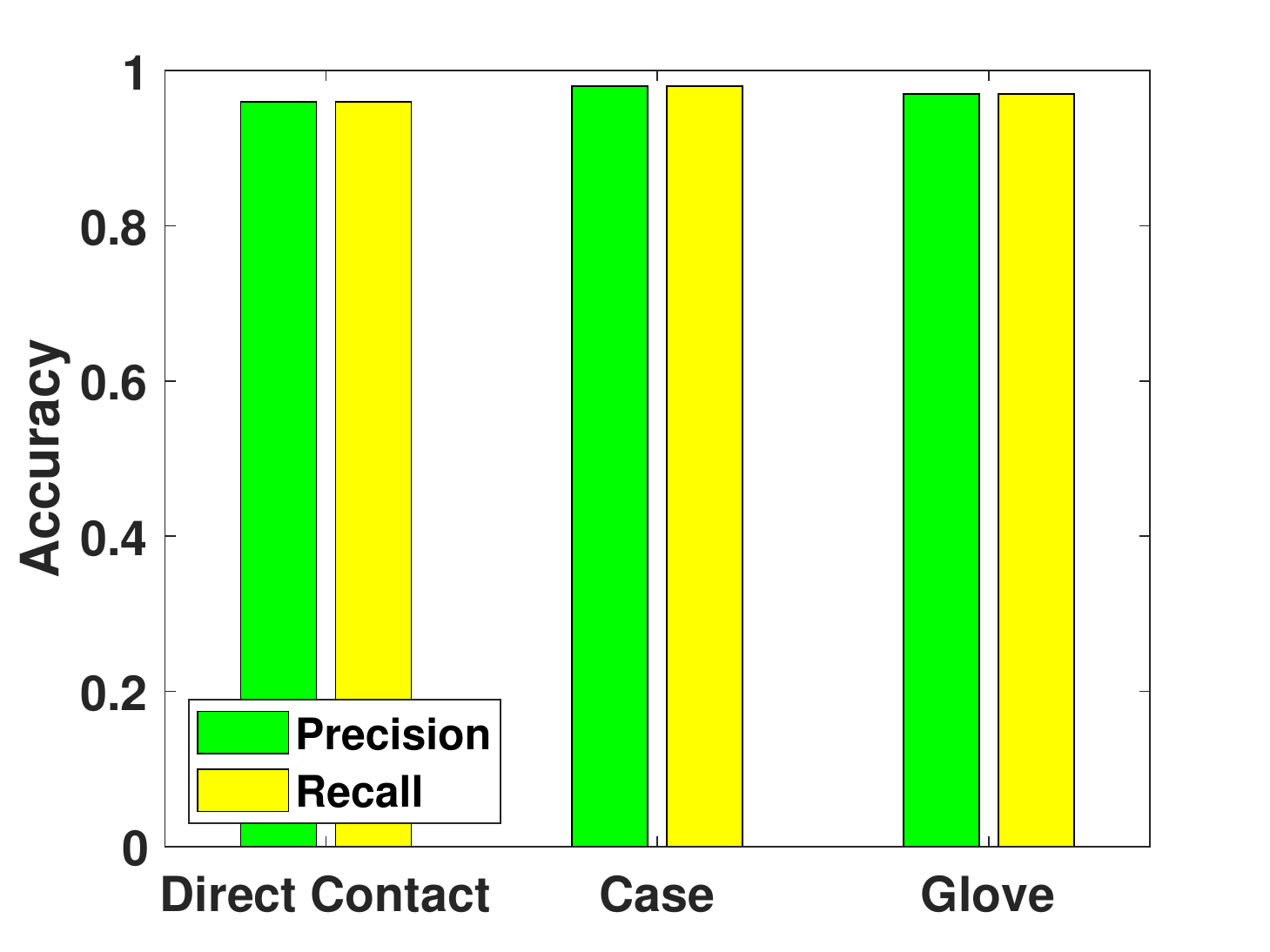}}%
	\hfill
	\vspace{-3mm}
	\caption{Performance when using the mobile device through direct vs. indirect physical contact. Results shown are for the Galaxy Note 5.}
	\vspace{-4mm}
	\label{fig:performance_factors}
\end{figure}



\textbf{Impact of Device Models.} 
We consider the performance discrepancies when operating \textit{EchoLock} on different mobile devices. Our participants are provided devices with our prototype installed and given a short explanation on its functionality. We note that a single demonstration less than 10 seconds long is sufficient for our participants to grasp how to operate \textit{EchoLock}, indicating its ease of use. We graph $P$ and $R$ in Figure~\ref{fig:performance_device} for the Nexus 5, Galaxy Note 5, and Tab A usage in office and public settings, and observe precision rates of 91\%, 90\%, and 81\%, respectively, when classified using our LDA classifier. When evaluated using our SVM classifier, we find precision rates of 94\%, 80\%, and 69\%, again respectively. Similar trends follow for recall rates. We note that performance can be correlated to the size of the device used, as users more easily acclimate to a favored posture for smaller devices compared to larger devices. We refer to these results, particularly for the Galaxy Note 5, as a benchmark of other experiments.


\textbf{Impact of Environmental Noises.} 
Figure \ref{fig:performance_enviro} shows performance differences when considering the surrounding environments during user authentication. The introduction of significant ambient noise produces a noticeable effect on accuracy, showing precision decline from 84\% to 78\% in the case of SVM and 92\% to 81\% for LDA. Our smartphone devices showed greater resilience under these conditions compared to our tablet device. We suspect this is attributed to the lack of secondary microphones on the Tab A device, which limits noise-cancellation capabilities compared to the smartphone devices. While interference from sources such as people walking or adjusting chairs caused an expected decrease in accuracy, we have found that loud vocalizations produced more significant performance degradation. We attribute this to our acoustic features, specifically our usage of MFCC. As MFCCs are most commonly used in speech processing, the presence of loud voices nearby causes our biometric measurements to be overwritten by the more dominant speech characteristics. 


\textbf{Indirect Physical Contact.} We also consider the influence of factors that may transform properties of either the device or user's hand when using \textit{EchoLock}. Figure~\ref{fig:performance_factors} shows our results for usage during situations where the user's hand does not directly make contact with the mobile device. To simulate these conditions, we equip our Galaxy Note 5 device with a smartphone case to change the physical properties of the medium. We also provide wool gloves roughly $2$mm thick for the user to wear during separate sets of experiments. We conduct our tests under office settings and compare to previously observed performance. Our findings do not show statistically significant deterioration for these conditions tested. On the contrary, some users showed slightly improved accuracy ranging from 1-3\%, particularly when wearing gloves. We suggest that the material of the gloves conformed well to the curvature of the hand while simultaneously suppressing certain variance in grip behavior, such as minor fidgeting or twitching. 

\begin{figure}[t]
	\centering
	\subfigure[Top microphone]{%
		\label{fig:mic_a}%
		\includegraphics[width=0.49\linewidth]{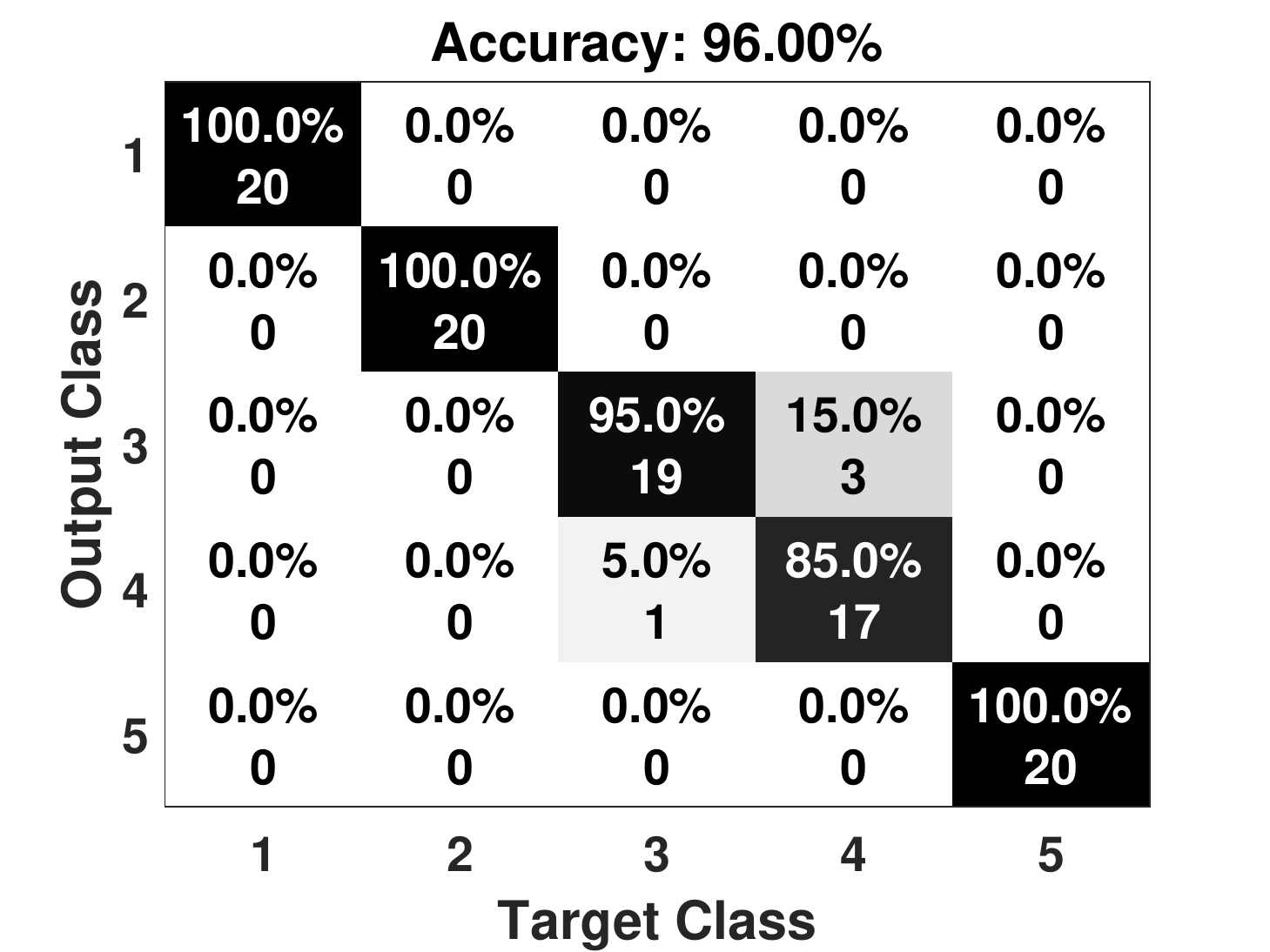}}%
	\hfill
	\subfigure[Bottom microphone]{%
		\label{fig:mic_b}%
		\includegraphics[width=0.49\linewidth]{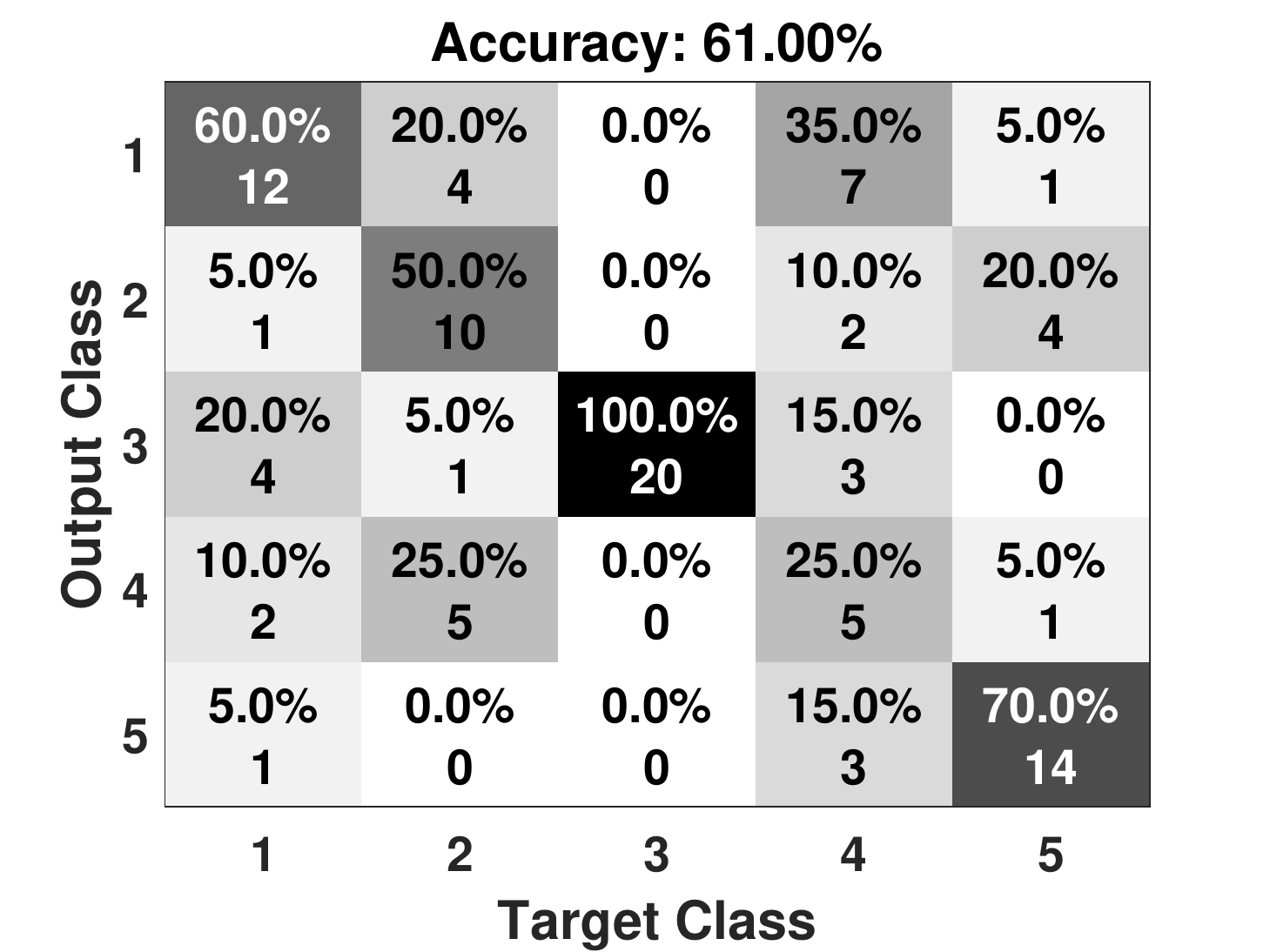}}%
	\hfill
	\vspace{-3mm}
	\caption{Classification accuracy for two available microphones based on signal transmission from a bottom-positioned speaker. }
	\vspace{-5mm}
	\label{fig:mics}
\end{figure}

\textbf{Impact of Embedded Microphone Location.}
For devices with more than a single onboard microphone, quality of identification can vary noticeably. Figure~\ref{fig:mics} shows an example of two microphones' ability to distinguish separate people based on features extracted from their recordings. Structure-borne sound originating from the bottom speaker will be considerably dampened by the force of the user squeezing the device when received by the bottom microphone. For the top microphone, this sound is more strongly influenced by the positioning of the thumb and fingers, which shapes the signal as it decays due to the longer distance traveled. When evaluated using $100$ samples provided from $5$ people, we observe identification accuracy of $96\%$ and $61\%$ for the top and bottom microphones, respectively. The comparatively less accurate rate for the bottom microphone is attributed to the shorter path from the source speaker to the receiving microphone, preventing sufficient vibration and feature formation. This stipulates that our speaker and microphone must be orientated further apart such that they encompass as much of the device as possible for optimal measurement. As such, we compute our results using data provided from speaker and microphone pairs with the greatest amount of separation when an option to select exists.



\begin{figure}[t]
	\centering
	\includegraphics[width=2 in]{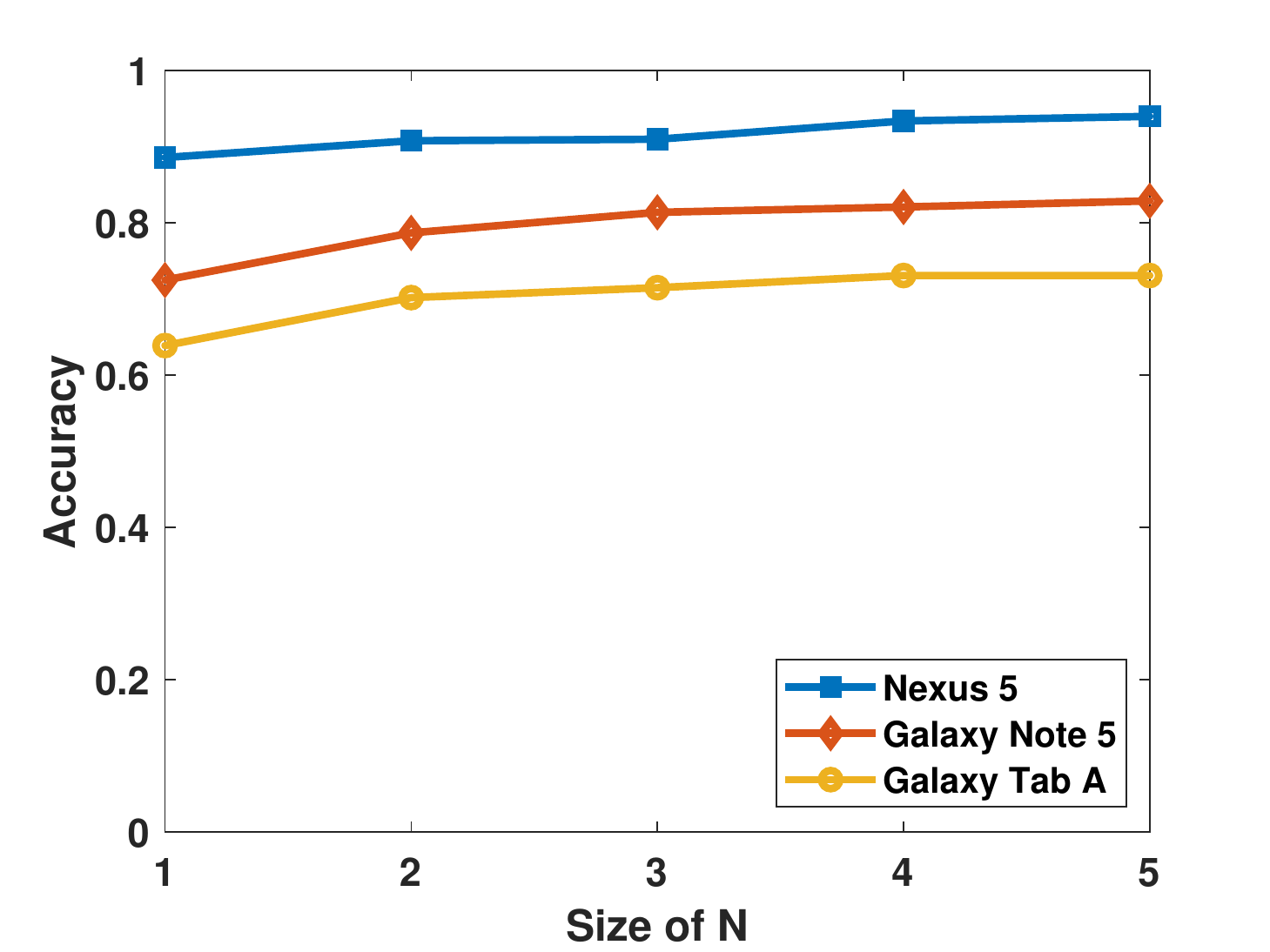}
	\vspace{-3mm}
	\caption{Performance of EchoLock as length of $n$-chirp sequence increases.}
	\vspace{-3mm}
	\label{fig:acc}
\end{figure}

\textbf{N-Chirp Sequence Length.} We investigate the effect of chirp sequence length on identification accuracy as a consideration to improve performance, shown in Figure~\ref{fig:acc}. While increasing the size of $n$ used for sensing naturally enhances accuracy, we observe an onset of diminishing returns rapidly after roughly 5 iterations. Stability around 90\% can be maintained for $n$ values as low as 3 in optimal test cases (i.e. smaller devices, quiet settings). As mentioned in Section~\ref{subsec:sigdesign}, a single chirp iteration requires only $25$ms, or $50$ms when accounting for buffering between iterations, meaning that our current performance can be feasibly achieved in execution times competitive with techniques such as fingerprinting. These results also indicate that an individual's hand biometrics are recognizable such that our classifiers may begin to identify them with moderate success using a relatively small number of training samples.


\subsection{Attacks on User Credentials}
\label{subsec:attack_eval}
\begin{figure}[t]
	\centering
	\subfigure[ROC curve for user identification aross 3 devices]{%
		\label{fig:roc}%
		\includegraphics[width=0.49\linewidth]{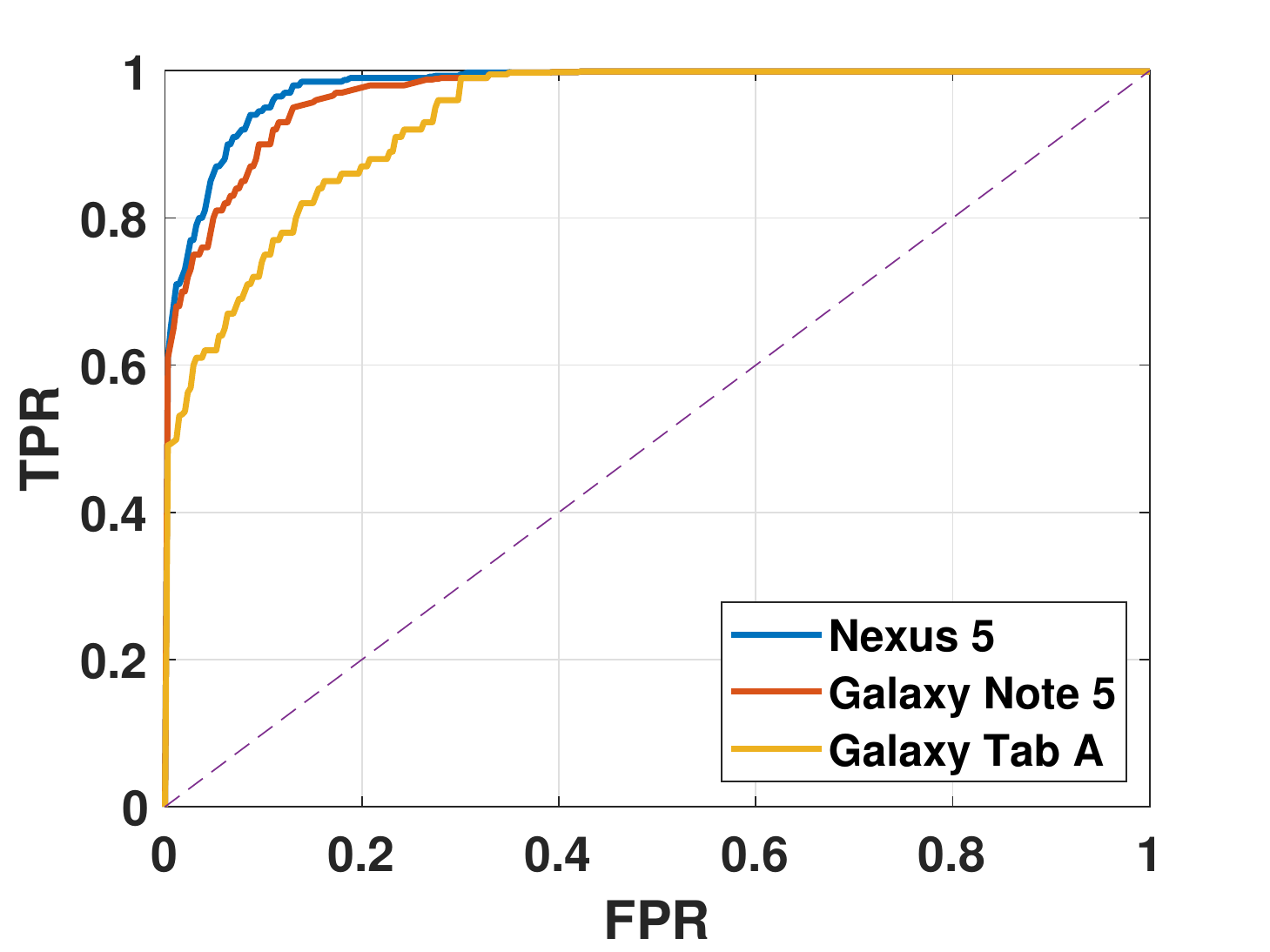}}%
	\hfill
	\subfigure[TP and FP for eavesdropping and replay attacks.]{%
		\label{fig:impersonate}%
		\includegraphics[width=0.49\linewidth]{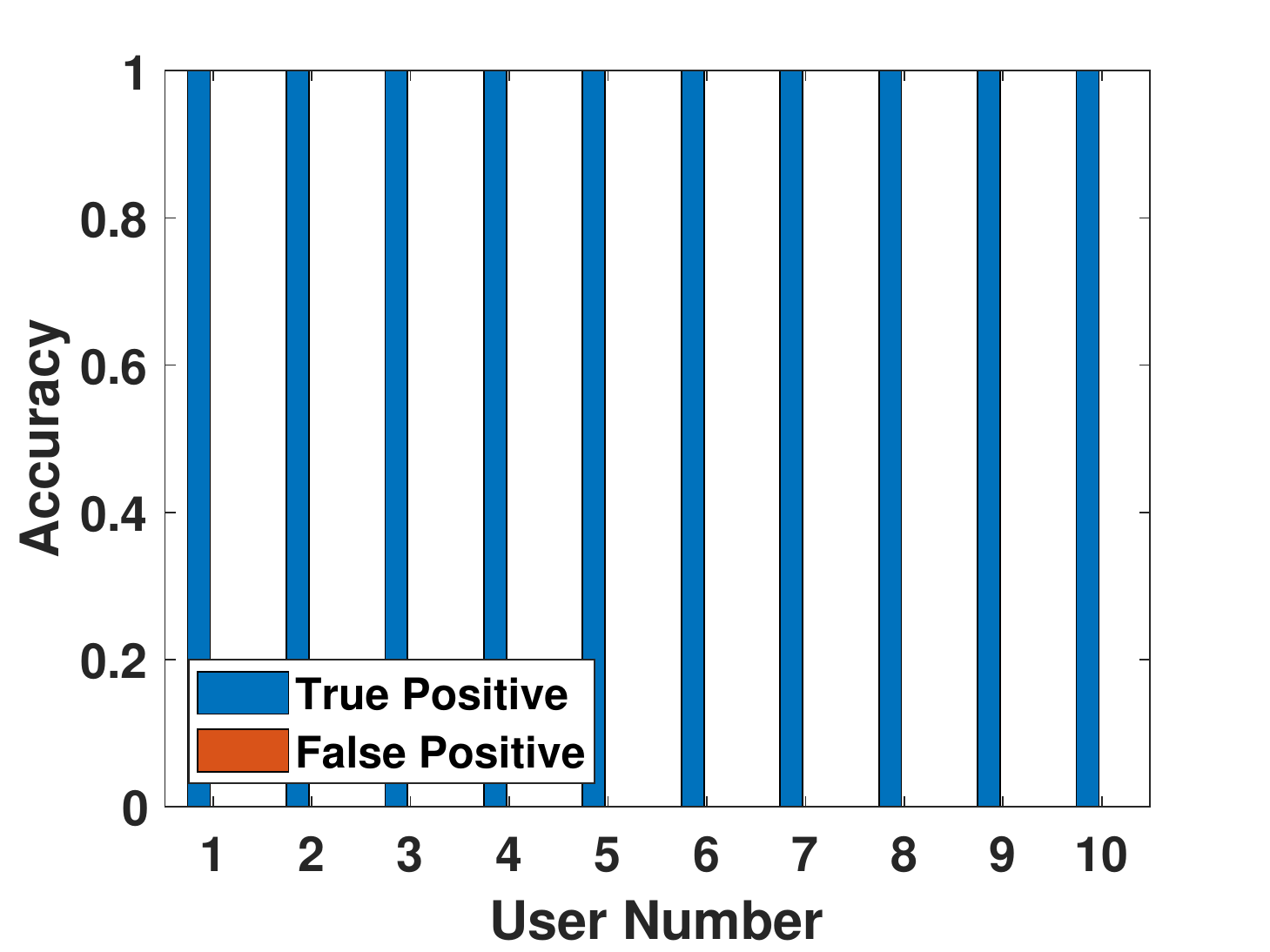}}%
	\vspace{-3mm}
	\caption{Performance of user identification during  evaluation of impersonation and eavesdropping/replay attacks.}
	\vspace{-3mm}
	\label{fig:attack}
\end{figure}

\textbf{Impersonation Attacks.} We evaluate the possibility of potential attackers to impersonate hand profiles of other users as a means of gaining unauthorized access to devices and information. For our assessment, we consider the worst case scenario; a limited (i.e. 5) training sample size for the victim user and multiple informed attackers. 20 participants first observe an instance of the designated victim using our hardware platforms and given 10 attempts to imitate their hold for each device. We measure the True Positive (TP) and False Positive (FP) rates and plot the ROC curve in Figure \ref{fig:roc}, showing FP rates as low as 6\% for a 90\% TP rate for devices such as the Nexus 5. These results suggest that observation alone is not sufficient to impersonate another hand profile; success is dependent on physical similarity between the attacker and victim, which the attacker cannot control.

\begin{figure}[t]
	\centering
	\includegraphics[width=\linewidth]{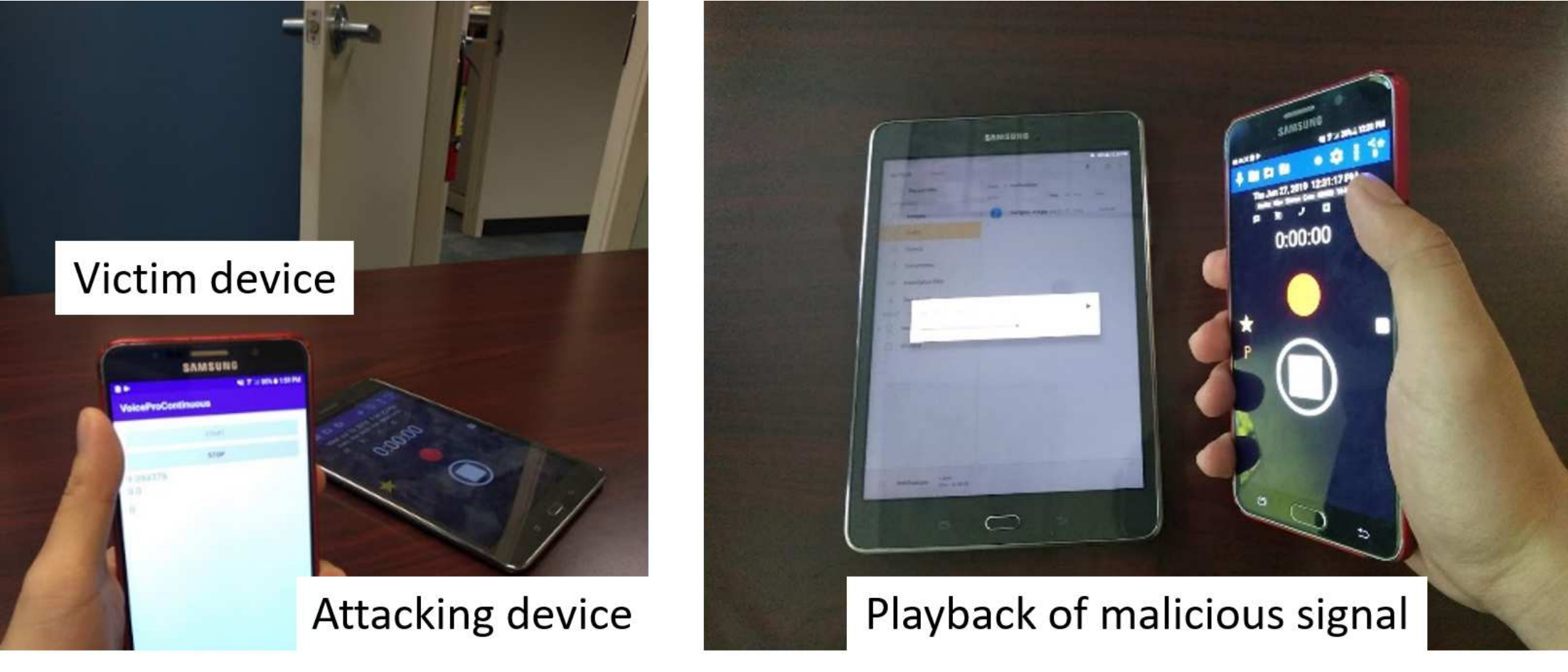}
	\caption{Example attack model setup.} 
	\vspace{-3mm}
	\label{fig:attack_setup}
\end{figure}

\begin{figure}[t]
	\centering
	\subfigure[Signal on victim device]{%
		\label{fig:eavesdrop_1}%
		\includegraphics[width=0.32\linewidth]{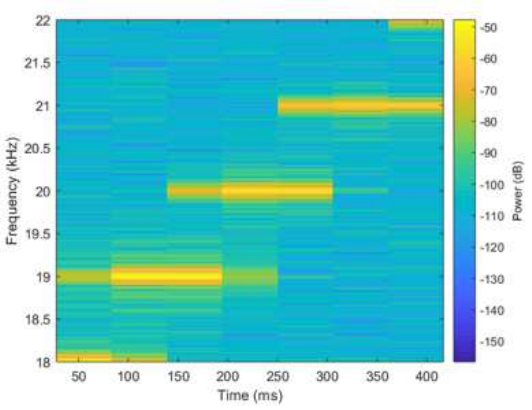}}%
	\hfill
	\subfigure[Eavesdropped signal]{%
		\label{fig:eavesdrop_2}%
		\includegraphics[width=0.32\linewidth]{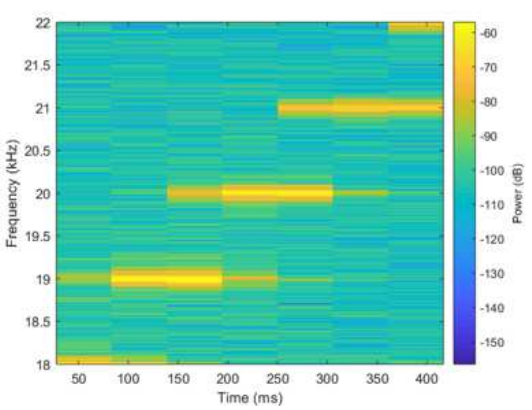}}%
	\vspace{-3mm}
	\subfigure[Replayed signal observed by victim]{%
		\label{fig:eavesdrop_3}%
		\includegraphics[width=0.32\linewidth]{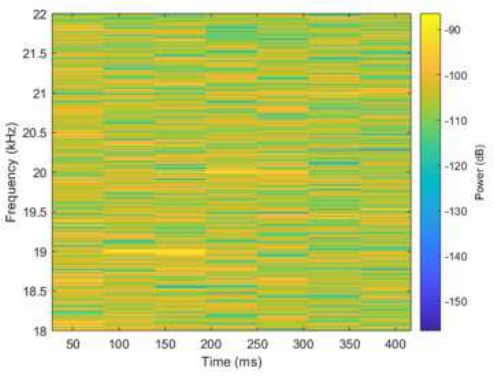}}%
	\hfill
	\caption{Frequency clarity throughout eavesdropping and replay attack 0.2m from victim.}
	\vspace{-5mm}
	\label{fig:eavesdrop}
\end{figure}

\textbf{Eavesdropping and Replay Attacks.} We assessed the viability of eavesdropping information by during our studies on standard usage behavior. Figure \ref{fig:attack_setup} shows an example configuration of hardware during these experiments. One of our mobile devices acts as a malicious sensor, listening for the validating signal used to identify the user. Filtered signals recovered from these simulated attacks in an office setting showed recognizable chirp patterns, as depicted in Figure \ref{fig:eavesdrop}, however clarity is lost due to the multipath effect. In trials with 10 users, 9 acting as victims and 1 as an attacker, directly replaying these recordings failed to grant the attacker access to the victim's device in all instances, depicted in Figure \ref{fig:impersonate}. Recorded signals must travel through two airborne paths, once from the victim device to the eavesdropper and vice versa, causing significant attenuation and loss of genuine structure-borne properties. Sophisticated attackers may attempt to apply noise cancellation and amplification to minimize attenuation, but careless processing of this signal risks destroying embedded features used to validate the user.

\section{Discussion}
\label{sec:discuss}


\textbf{Optimal Scenarios for Classifiers.} Our findings show situational advantages for employing SVM and LDA classifiers for user identification. Figures \ref{fig:performance_device} and \ref{fig:performance_enviro} indicate that both the device model and surrounding environments contribute significantly to the performance of \textit{EchoLock}. These results indicate that our LDA classifier is more platform-agnostic, showing less than 1\% precision and recall difference for different smartphones and under 10\% difference between smartphones and tablets. In contrast, we note that SVM is less susceptible to change in performance as a result of additional background noise, showing an average 6\% difference between office and public spaces compared to 11\% for LDA. These findings are intuitive to the nature of their respective classifiers as the linearity of LDA is suited for situations with consistency (i.e. a quiet space with devices and hand geometry that do not change) whereas the introduction of unpredictable factors requires adaptability offered by techniques such as SVM.

\begin{figure}[t]
	\centering
	\subfigure[Frequency domain under ordinary usage]{%
		\label{fig:jam_1}%
		\includegraphics[width=0.49\linewidth]{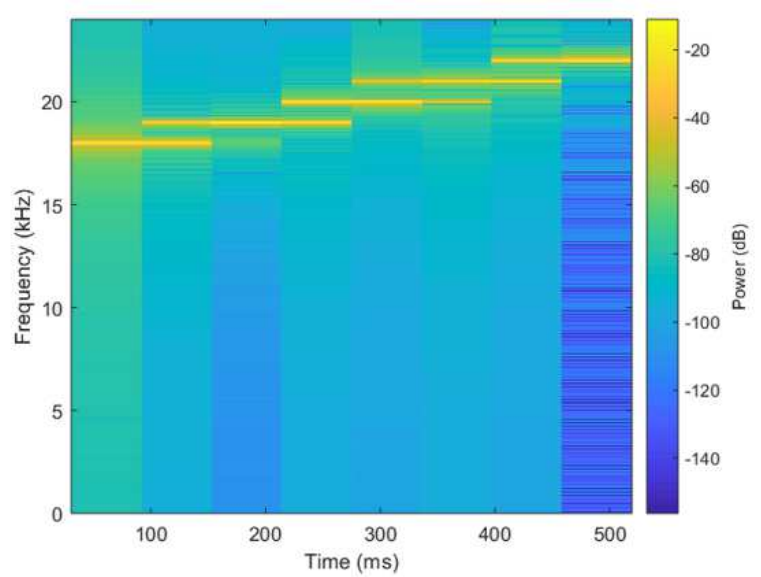}}%
	\hfill
	\subfigure[Frequency domain under jamming attack]{%
		\label{fig:jam_2}%
		\includegraphics[width=0.49\linewidth]{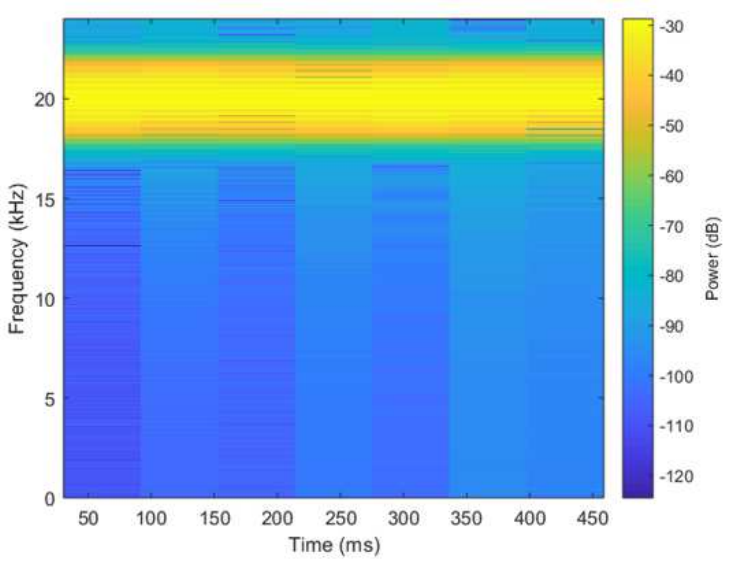}}%
	\vspace{-3mm}
	\caption{Identification of jamming attempts within the system operational frequency range.}
	\vspace{-5mm}
	\label{fig:jamming}
\end{figure}

\textbf{Jamming Attacks.} We also consider the possibility of acoustic disruptions to our performance via jamming strategies. To do so, we study the ability for 10 users to use our system when an attacking device plays a continuous signal within the operational frequency of \textit{EchoLock}. Figure \ref{fig:jamming} shows examples of the frequency domain during these experiments. We play an attacking signal continuously sweeping from 18kHz to 22kHz during ten identification attempts per user at a distance of $0.2$m. We find that detection of these jamming signals is feasible using a threshold scheme. We note that although detection is possible, negating the interference still poses a challenge. However, jamming attempts from distances greater than 1m were observed to lose potency, functioning more similar to public environment conditions described in Section \ref{subsec:impacts}. Methods to avoid threshold detection may attempt to limit the extent they exceed normal intensity, however this requires careful synchronization on the scale of milliseconds. Additionally, the jamming signal must match the length of the $n$-chirp sequence, which the attacker cannot predict. We are further improving resistance to these attacks as part of our future work.




\textbf{Potential Hardware Constraints.} 
During our selection of candidate devices to experiment on, we became aware of certain hardware configurations unsuitable to implement \textit{EchoLock} on (e.g. speakers on front face, microphone on back). We determine that our implementation necessitates our hardware components to be orientated such that they are as distant from each other as possible to allow for uninterrupted sound propagation, which cannot be accurately measured with a backward-facing microphone. Adherence to this may be somewhat relaxed by leveraging gyroscopic measurements to compensate for inconvenient speaker or microphone placement, though this also adds additional hardware requirements to an intentionally minimalist design. 


%


\section{Conclusion}
We have proposed \textit{EchoLock}, an inexpensive, non-intrusive, and lightweight identification protocol deployable on commodity mobile devices or smart IoT devices. Our system validates the user's identity using a novel technique leveraging acoustic sensing of structure-borne sound propagation to ascertain biometric characteristics based on how the user holds their devices. A prototype of \textit{EchoLock} has been implemented on multiple Android platforms and evaluated in key use case scenarios. Our identification technique is quick to conduct, low effort to use, and demonstrates accuracy over $90\%$. 
Our current prototype evaluates \textit{EchoLock} as a standalone identification protocol; for future work, we intend to integrate it with existing authentication techniques and assess the possibility of elevating current security rates. Furthermore, we intend to also enhance our ability to defend against more sophisticated attack models. While we have demonstrated \textit{EchoLock} on select mobile devices, we are also exploring other hardware options and developing a more robust implementation to realize high accuracy, low effort identification. 
\vspace{-1mm}

	%
	%
	%
	%
	
	%
	%
	
	
	
	%
	\bibliographystyle{acm}
	\bibliography{../bib/main}  
	%
	%
	

\appendix
\label{sec:appendix}
\begin{figure}[h]
	\centering
	\subfigure[Glove equipped]{%
		\label{fig:glove}%
		\includegraphics[height= 1.5 in]{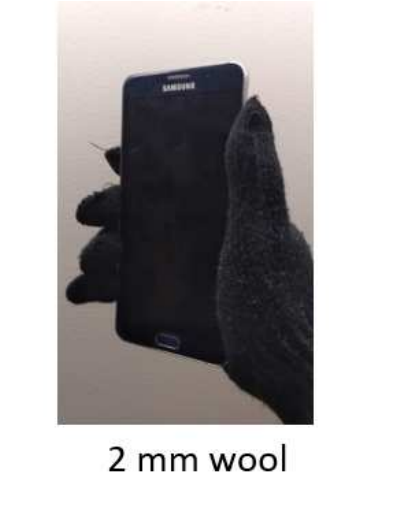}}
	\subfigure[Protective case]{%
		\label{fig:case}%
		\includegraphics[height= 1.5 in]{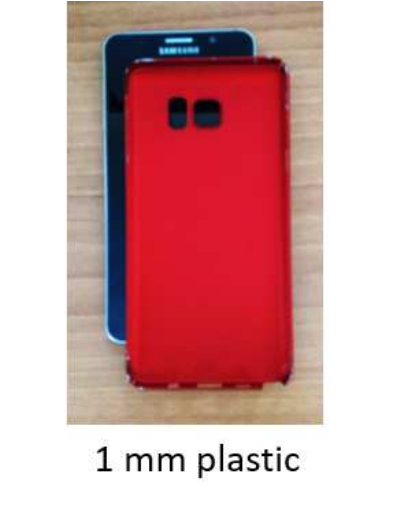}}
	\hfill
	\caption{Considerations for device usage through indirect physical contact. Protective case is detached for visual clarity.}
	\vspace{-5mm}
	\label{fig:indirect_use}
\end{figure}

\end{document}